\documentclass[11pt]{article}
\pdfoutput=1
\usepackage{amsmath, amsfonts, amssymb}
\usepackage{comment}
\usepackage{graphicx}
\usepackage{psfrag}
\usepackage{amsthm}
\usepackage[usenames,dvipsnames,svgnames,table]{xcolor}
\usepackage{enumerate}
\usepackage{tcolorbox}
\usepackage{mathtools}
\usepackage{soul}
\usepackage{subfigure}
\usepackage{mathrsfs}  
\usepackage{a4wide}
\usepackage{booktabs}
\usepackage{tikz}
\usepackage{tikz-cd}
\usepackage{makecell}
\usepackage{multicol}
\usepackage{braket}
\usepackage{bbold}

\usepackage{color}
\definecolor{dark-gray}{gray}{0.20}
\definecolor{gray}{gray}{0.30}
\definecolor{light-gray}{gray}{0.80}
\definecolor{dark-red}{rgb}{0.7,0,0}
\definecolor{dark-green}{rgb}{0.1,0.4,0}
\definecolor{dark-blue}{rgb}{0.3,0.3,0.7}
\definecolor{light-blue}{rgb}{0.8,0.8,1}
\definecolor{swamp}{RGB}{240, 199, 197}

\usepackage{pifont}
\usepackage{setspace}

\newcommand{\be}{\begin{equation}}
\newcommand{\ee}{\end{equation}}

\def\be{\begin{equation}}
\def\ee{\end{equation}}
\def\bea{\begin{eqnarray}}
\def\eea{\end{eqnarray}}

\newcommand{\arXivold}[2]{\href{http://arxiv.org/pdf/#1}{{\tt #2/#1}}}

\def\simleq{\; \raise0.3ex\hbox{$<$\kern-0.75em
		\raise-1.1ex\hbox{$\sim$}}\; }
\def\simgeq{\; \raise0.3ex\hbox{$>$\kern-0.75em
		\raise-1.1ex\hbox{$\sim$}}\; }

\numberwithin{equation}{section}

\usepackage{jheppub} 
\hypersetup{
	colorlinks=true,
	linkcolor=dark-blue,
	citecolor=dark-red,
	urlcolor=dark-green,
	linktoc=page,
	pageanchor=false
}

\title{\centering Resonantly-Enhanced Baryogenesis through Asymmetric Capture by Primordial Black Holes}

\author{Fayez Abu-Ajamieh$^1$,}
\affiliation{$^1$Formerly: The Indian Institute of Science, Bangalore, India}
\emailAdd{fayezabuajamieh@gmail.com}

\author{Xiaoyan Huo$^2$,}
\affiliation{$^2$Universiti Teknologi Malaysia, Malaysia}
\emailAdd{huoxiaoyanphd@163.com}

\abstract{Primordial black holes (PBHs) can generate the Baryon Asymmetry of the Universe (BAU) through asymmetric capture of baryon-charge-carrying particles, but previous realizations of this mechanism typically yield an asymmetry well below the observed value because the relevant CP-violating scattering asymmetry is loop suppressed. We show that this limitation can be overcome through resonant CP violation in a mixed system of two nearly degenerate unstable fermions, $Y_{1,2}$, whose mass splitting is comparable to their decay widths. Rare production of the coherent $Y_1$--$Y_2$ state, followed by CP-asymmetric return and loss decays, generates an $\mathcal{O}(0.1)$ difference in integrated decay probabilities while preserving the global baryon charge at the Lagrangian level. Embedding this resonant source in the PBH-capture transport system yields the observed baryon asymmetry. We first reproduce the mechanism in a zero-temperature, sudden-evaporation treatment, which gives $\eta_B^{T=0}\simeq1.74\times10^{-10} \simeq2.00\eta_B^{\rm obs}$. We find that after including thermal corrections and continuous PBH evaporation and entropy production, the resulting BAU becomes $\eta_B = 8.72\times 10^{-11}$ in agreement with the observed value. An interesting finding of the model is that even if PBH absorption remains active, we would still have BAU $\eta_{B}^{\text{no evap}} \sim 9.1 \times 10^{-12} \sim 0.105 \eta_B^{\rm obs}$. We further investigate gravitational-wave signatures associated with the primordial fluctuations responsible for PBH formation. The corresponding scalar-induced gravitational-wave spectra peak at $f \sim 0.16\text{--}0.17~\mathrm{MHz}$, with $h^2\Omega_{\rm GW}^{\rm peak} \sim (3\text{--}7)\times 10^{-8}$. Under idealized sensitivity assumptions, U-DECIGO can achieve signal-to-noise ratios above 10 for the benchmark spectra.}

\begin{document}

\makeatletter
\let\old@fpheader\@fpheader

\makeatother

\maketitle

\section{Introduction}\label{intro}
The fact that the universe has more baryons than antibaryon, commonly referred to as the baryon asymmetry of the Universe (BAU), remains one of the main open problems in physics today. This asymmetry can be expressed in the measured quantity~\cite{Planck:2018vyg, ParticleDataGroup:2020ssz, Schoneberg:2024ifp}\footnote{BAU is also commonly expressed as the baryon-to-photon ratio $\eta = \frac{n_{B}-n_{\overline{B}}}{n_{\gamma}}\simeq 6\times 10^{-10}$.}
\begin{equation}\label{eq:BAU_value}
\eta_{B} \equiv \frac{n_{B}-n_{\bar{B}}}{s} \simeq 8.7 \times 10^{-11}\,,
\end{equation}
where $n_B$ ($n_{\bar B}$) is the number density of baryons (antibaryons) and $s$ is the entropy density. According to the Sakharov criteria~\cite{Sakharov:1967dj}, a successful BAU requires the following ingredients:
\begin{enumerate}
\item C and CP violation,
\item Baryon number violation\footnote{In the present scenario, the particle-sector interactions conserve the global $U(1)_B$ charge; the visible-sector baryon asymmetry instead arises because PBHs selectively sequester one sign of baryon charge and subsequently evaporate, leaving an unequal baryon charge in the surviving plasma.}, and
\item A departure from thermal equilibrium.
\end{enumerate}

Although the Standard Model (SM) contains the ingredients relevant to the Sakharov criteria, they are insufficient to account for the observed BAU~\cite{Bochkarev:1987wf, Kajantie:1995kf, Huet:1994jb, Gavela:1994dt}, which calls for physics beyond the SM (BSM). There  are many proposed solutions to the BAU, including GUT baryogenesis~\cite{Yoshimura:1978ex}, baryogenesis through leptogenesis~\cite{Fukugita:1986hr}, Electroweak (EW) baryogenesis~\cite{
Cohen:1993nk, Rubakov:1996vz, Morrissey:2012db, vandeVis:2025efm}, Affleck–Dine baryogenesis~\cite{Affleck:1984fy}, and gravitational baryogenesis~\cite{Davoudiasl:2004gf}.

In~\cite{Dolgov:2020kqj,Dolgov:2021tsa}, a novel mechanism for producing baryon asymmetry through the asymmetric capture of baryons and antibaryons by PBHs~\cite{Khlopov:2008qy, Belotsky:2014kca, Abu-Ajamieh:2024gaw, Abu-Ajamieh:2024xic, Abu-Ajamieh:2024egb, Abu-Ajamieh:2025sul} was proposed. Specifically, the mechanism assumes the existence of a baryon-charge-carrying particle $X$ of mass $m_{X}$ and its antiparticle $\overline{X}$. The interaction of $X$ and $\overline{X}$ with the plasma surrounding PBHs is assumed to violate C and CP (but not baryon number), which leads to differences in their mobility towards the PBH, resulting in an asymmetry between the capture of $X$ and $\overline{X}$ thereby creating the observed BAU.

More careful analysis of this mechanism in~\cite{Ambrosone:2021lsx}, however, revealed two major issues: first, continued PBH absorption would eventually deplete the $X$ and $\overline X$ populations together with the generated asymmetry, unless PBH evaporation terminates the absorption sufficiently early; and second, careful evaluation showed that the CP-violation efficiency was significantly smaller than the temperature-independent value assumed in the earlier phenomenological treatment. To remedy this situation, Ref.~\cite{Ambrosone:2021lsx} proposed that PBHs should evaporate before they could erase the accumulated BAU. A detailed evaluation via the Boltzmann equations showed that such a proposal would indeed be successful at creating a BAU, however, it was shown that the resulting BAU is too small, reaching only $\eta_B\sim10^{-13}$, about three orders of magnitude below the observed value.

The reason behind this difficulty is that a successful creation of the observed BAU would require a significant CP asymmetry in the scattering cross sections of $X$ and $\overline{X}$, however, such an asymmetry is loop-generated and is suppressed at low temperature by the efficiency factor $\epsilon(x) \propto (m_{X}/T)^{-3/2}$, and when absorption by PBHs becomes significant, the CP asymmetry becomes small, leading to a maximum asymmetry of $\eta_{B} \sim 10^{-13}$.

In this paper, we present a successful model for creating the observed BAU through resonantly enhanced asymmetric capture by PBHs. The idea of producing significant CP violation through resonant enhancement is not new~\cite{Pilaftsis:1997dr}, and has recently been applied to BAU~\cite{Kilic:2026ach}; however, here we utilize it to build a model that can successfully produce the observed BAU through the asymmetric capture by PBHs. The general idea of the resonant enhancement of CP violation goes as follows: consider two nearly degenerate unstable particles whose mass splitting is comparable to their decay widths $\Delta M \sim \Gamma$. These two intermediate particles interfere and the off-diagonal self-energies generate CP-asymmetry that can be $\mathcal{O}(1)$.

We show that through resonant enhancement, the mechanism of asymmetric capture by PBHs can successfully produce the observed BAU. We first present a treatment similar to~\cite{Ambrosone:2021lsx}, then we improve upon it by incorporating finite-temperature effects, continuous PBH evaporation, and continuous radiation and entropy evolution. In addition, we explore the detection prospects of the Scalar-Induced Gravitational Wave (SIGW) background associated with this mechanism.

This paper is organized as follows: in Section~\ref{sec:Model} we present the model and the asymmetry production channels and discuss the resonant enhancement. In Section~\ref{sec:BAU_T0}  we present the mechanism and calculate the resulting BAU neglecting thermal corrections and assuming an abrupt decay of PBHs. In Section~\ref{sec:thermal} we redo the calculations after incorporating the thermal corrections and treating the evaporation of PBHs as a continuous process. In Section~\ref{sec:GW} we discuss the SIGWs associated with PBH formation and their detection prospects in future experiments, and finally, we conclude in Section~\ref{sec:Discussion}. We relegate many of the technical details to the appendices.

\section{Model and main mechanism}
\label{sec:Model}
\subsection{Lagrangian and field content}

To implement our mechanism, we propose a generalized model similar to Ref.~\cite{Ambrosone:2021lsx}. However, instead of a single fermion $Y$, we introduce two nearly degenerate unstable Dirac fermions $Y_1$ and $Y_2$, whose mass splitting is comparable to their decay widths. In addition, we introduce three fermionic loss states $F_r$ and three scalar loss states $S_r$, with $r=1,2,3$. The interaction Lagrangian is given by
\begin{equation}\label{eq:int_lag}
\mathcal{L} \supset -g_{aX}\phi\overline{a}X - g_{cX} \phi\overline{c}X - \sum_{i=1}^{2} g_{bY_{i}}\phi \overline{b}Y_{i} - \sum_{i=1}^{2} y_{i}\psi \overline{Y}_{i}X - g_{ba}\psi \overline{b}a - \sum_{i=1}^{2}\sum_{r=1}^{3}\kappa_{ir}S_{r}\overline{F}_{r}Y_{i} + \mathrm{h.c.}\,,
\end{equation}
where $X$ is a heavy baryon-charge-carrying particle, $a$, $b$ and $c$ are bath species, $\phi$  mediates the rare production process, while $\psi$ mediates the return decay and its bath interaction. The global $U(1)_{B}$ charge assignments are given in Table~\ref{tab:U1B_charges}. Notice that the particle-sector Lagrangian preserves the global $U(1)_B$ charge. We assume additional baryon-conserving interactions of $F_r$ and $S_r$ with the SM that keep the loss-sector daughters thermally populated over the relevant temperature range and eventually transfer the spectator baryon charge to visible baryons. Thus, the produced BAU does not arise from explicit baryon-number violation in the particle-sector interactions, but from the asymmetric sequestration of charge by PBHs.  For concreteness, we use the following benchmark values throughout this paper:
\begin{equation}
\begin{array}{lll}
m_X = 130~\mathrm{GeV}\,,
&
M_{Y_1} = 170~\mathrm{GeV}\,,
&
\Delta M = M_{Y_2}-M_{Y_1}=16.593~\mathrm{MeV}\,,
\\[1mm]

m_b = 10~\mathrm{GeV}\,,
&
m_\psi = 5~\mathrm{GeV}\,,
&
m_\phi = 179~\mathrm{GeV}\,,
\\[1mm]

\kappa_1 = \kappa(1,1,1)\,,
&
\kappa_2 = \kappa(1,e^{i\phi_\kappa},e^{2i\phi_\kappa})\,,
&
\kappa = 0.04\,,\qquad \phi_\kappa = 1.03358\,,
\\[1mm]

y_1 = 0.04\,,
&
y_2 = 0.04e^{i\theta_y}\,,
&
\theta_y = 2.16421\,,
\\[1mm]
g_{aX} = f = 7\times10^{-5}\,,
&
g_{bY_1} = g_{bY_2}\equiv g_b = 1.281\times10^{-5}\,,
& 
h= 3.29 \times 10^{-10}\,.
\end{array}
\label{eq:benchmark}
\end{equation}
We take $m_{F_r},m_{S_r}\ll M_Y$ and neglect their masses in the loss-channel phase space.
\begin{table}[t!]
\centering
\begin{tabular}{c|ccccccc}
\hline
Field & $X$ & $Y_1,Y_2$ & $F_r$ & $a,b,c$ & $\phi$ & $\psi$ & $S_r$ \\
\hline
$U(1)_B$ & $+1$ & $+1$ & $+1$ & $0$ & $-1$ & $0$ & $0$ \\
\hline
\end{tabular}
\caption{Global $U(1)_B$ charge assignments.}
\label{tab:U1B_charges}
\end{table}

\subsection{Rare production and resonance}
\begin{figure}[!t]
    \centering
    \includegraphics[width=0.8\textwidth]{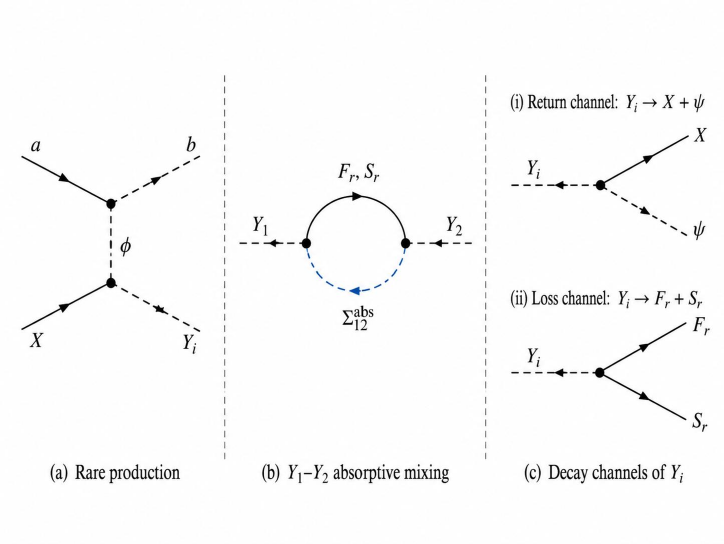}
    \caption{Resonant CP source. (Left) rare production, (middle) mixed propagation, and (right) return and loss decay channels. Here $\Sigma_{12}^{\mathrm{abs}}$ denotes the absorptive self-energy.}
    \label{fig:feynman}
\end{figure}

The resonant mechanism starts with the rare production process
\begin{equation}\label{eq:process1}
X + \overline{a} \rightarrow Y_{i} + \overline{b}\,,
\end{equation}
which coherently produces $Y_{1}$ and $Y_{2}$. These nearly degenerate states undergo coherent $Y_1$--$Y_2$ mixing and propagate as a mixed unstable system before decaying. We assume that the production process does not resolve the small mass splitting $\Delta M$, so that the produced state is a coherent
superposition of $Y_1$ and $Y_2$ rather than an incoherent mixture. There are two classes of decay channels. The first process which we call the return channel,
\begin{equation}\label{eq:process2}
Y_{i} \rightarrow X + \psi\,,
\end{equation}
returns the baryon charge to the $X$ sector, whereas the other, which we call the loss channel, 
\begin{equation}\label{eq:process3}
Y_{i} \rightarrow F_{r} + S_{r}\,,
\end{equation}
transfers the baryon charge from the $X$ sector to the spectator sector $F_{r}$. As we will show below, the interference between $Y_{1}$ and $Y_{2}$ makes the probabilities of these processes different from their CP-conjugate processes, thereby creating asymmetric mobility of $X$ and $\overline{X}$ that creates the observed BAU. The Feynman diagrams of this sequence are shown in Figure~\ref{fig:feynman}. Notice that, unlike Ref.~\cite{Ambrosone:2021lsx}, the dominant CP-violating contribution arises from resonantly enhanced $Y_1$--$Y_2$ mixing rather than from the interference between the tree-level and one-loop amplitudes shown in Figure~\ref{fig:OriginalProcesses}. The latter nonresonant contribution is still present, but for the benchmark considered here it is many orders of magnitude smaller than the resonant source, as shown explicitly in Appendix~\ref{app:D}.

For the production process in Eq.~(\ref{eq:process1}), the relevant interactions are $g_{aX}\phi\bar a X$ and $g_{bY_i}\phi\bar bY_i$. The process therefore proceeds through $s$-channel $\phi$ exchange, with amplitude
\begin{equation}
\label{eq:matrix1}
\mathcal{M}_i^P
=
\frac{g_{aX}g_{bY_i}^{*}}
{s-m_\phi^2}
[\bar v_a(q)u_X(p)]
[\bar u_{Y_i}(k)v_b(\ell)]\,.
\end{equation}
After averaging over the initial spins and summing over the final spins,
\begin{equation}
\label{eq:matrix2}
\overline{|\mathcal M_P|^2}
=
4|g_{aX}|^2
\sum_{i=1}^2 |g_{bY_i}|^2
\frac{(p\cdot q)
      (k\cdot\ell-M_Ym_b)}
     {(s-m_\phi^2)^2}\,.
\end{equation}
For a massless bath fermion $a$,
\begin{equation}
p\cdot q=\frac{s-m_X^2}{2},
\qquad
k\cdot\ell=\frac{s-M_Y^2-m_b^2}{2}.
\end{equation}
We again take $M_Y\simeq M_{Y_1}\simeq M_{Y_2}$ in the production kinematics, since $\Delta M\ll M_{Y_{1,2}}$, while retaining the splitting in the coherent propagation. The production cross section is therefore
\begin{equation}
\label{eq:Xsection}
\sigma_P(s)
=
\frac{1}{16\pi(s-m_X^2)^2}
\int_{t_-}^{t_+}dt\,
\overline{|\mathcal M_P|^2}\,,
\end{equation}
where $t_\pm$ are given in Appendix~\ref{app:A}. 

\subsection{Mixed unstable system and integrated probabilities}
The decay widths of the loss and return processes in Eqs.~(\ref{eq:process2}) and~(\ref{eq:process3}), which are shown in the right Feynman diagrams of Figure~\ref{fig:feynman}, are given by
\begin{align}
(\Gamma^{L})_{ij}
&=
\frac{M_Y}{16\pi}
\sum_{r=1}^{3}
\kappa_{ir}\kappa_{jr}^{*}\,,
&
(\Gamma^{R})_{ij}
&=
\frac{M_Y}{16\pi}
\Phi_R(M_Y,m_X,m_\psi)\,
y_i y_j^{*}\,,
\end{align}
where we neglect the masses of the loss states $F_r$ and $S_r$ relative to $M_Y$, and
\begin{equation}
\label{eq:Phi}
\Phi_R(M_Y,m_X,m_\psi)
=
\frac{
\lambda^{1/2}(M_Y^2,m_X^2,m_\psi^2)
\left[(M_Y+m_X)^2-m_\psi^2\right]
}{
M_Y^4
}\,.
\end{equation}
Here and below we neglect the small mass splitting in the decay kinematics and take
$M_Y\simeq M_{Y_1}\simeq M_{Y_2}$, while retaining $\Delta M$ in the coherent propagation. The effective Hamiltonian can be expressed as
\begin{equation}
\label{eq:Hamiltonian}
H
=
M-\frac{i}{2}\left(\Gamma^L+\Gamma^R\right),
\qquad
M=
\begin{pmatrix}
M_{Y_1} & 0\\
0 & M_{Y_2}
\end{pmatrix}.
\end{equation}

Let the normalized initial production vector of the $Y_i$ system be
$p=(p_1,p_2)^T$. Since the rare production process is controlled by the couplings $g_{bY_i}$, the corresponding normalized production vector is
\begin{equation}
\label{eq:productionvector}
p_i
=
\frac{g_{bY_i}^{*}}
{\sqrt{|g_{bY_1}|^2+|g_{bY_2}|^2}}\,,
\qquad
p^\dagger p=1.
\end{equation}
For the benchmark in Eq.~(\ref{eq:benchmark}), where $g_{bY_1}=g_{bY_2}=g_b$ are real, this becomes
\begin{equation}
p
=
\frac{1}{\sqrt{2}}
\begin{pmatrix}
1\\
1
\end{pmatrix}.
\end{equation}

\begin{figure}[!t]
\centering
\includegraphics[width=0.7\textwidth]{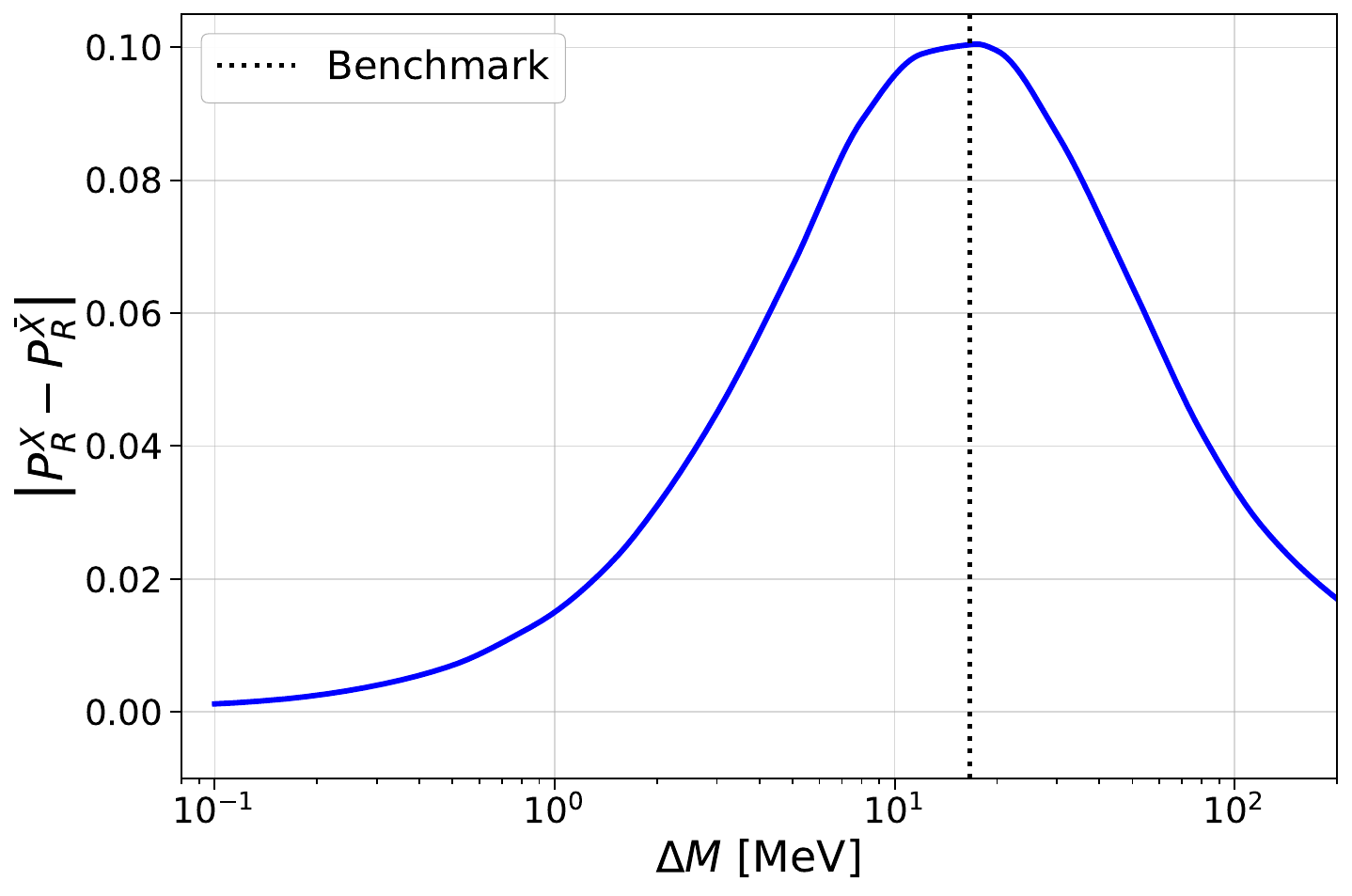}
\caption{Zero-temperature resonant return-probability difference as a function of the mass splitting of $Y_1$ and $Y_2$. The benchmark point lies near the maximum of $|\Delta P_R|$.}
\label{fig:ReturnProb}
\end{figure}

The corresponding probabilities for the CP-conjugate process are obtained by complex conjugating the CP-violating couplings in the production and decay matrices. For the present benchmark the production vector is real and therefore unchanged under CP, while the complex phases in $y_i$ and $\kappa_{ir}$ are conjugated. Thus the probability of ending in channel $c=L,R$ is given by
\begin{equation}
\label{eq:Pc}
P_c
=
\int_0^\infty dt\,
\mathrm{Tr}\!\left[\Gamma_c\rho(t)\right]
=
p^\dagger K_c p\,,
\end{equation}
where the density matrix is
\begin{equation}
\rho(t)
=
e^{-iHt}
pp^\dagger
e^{iH^\dagger t}\,,
\end{equation}
and $K_c$ is obtained from the Sylvester equation (see Appendix~\ref{app:B} for the derivation)
\begin{equation}
\label{eq:Sylvester}
H^\dagger K_c-K_cH=i\Gamma_c\,.
\end{equation}
With this prescription, we resum coherent oscillations, damping, and interference. Notice that since
$\Gamma=\Gamma_L+\Gamma_R$, one has $K_L+K_R=\mathbb{1}$ and consequently $P_L+P_R=1$, as required. For the benchmark in Eq.~(\ref{eq:benchmark}), we find
\begin{align}
P_L^X
&\simeq
0.76720\,,
&
P_R^X
&\simeq
0.23280\,,
\\
P_L^{\overline X}
&\simeq
0.86760\,,
&
P_R^{\overline X}
&\simeq
0.13240\,.
\end{align}
Thus,
\begin{equation}
\Delta P_R
\equiv
P_R^X-P_R^{\overline X}
\simeq
0.1\,,
\qquad
\Delta P_L
=
-\Delta P_R\,.
\end{equation}
The relation $\Delta P_L=-\Delta P_R$ follows directly from probability conservation and CPT invariant,
$P_L+P_R=1$, for both the particle and CP-conjugate systems. Figure~\ref{fig:ReturnProb} shows the zero-temperature resonant return-probability difference as a function of the mass splitting. The asymmetry reaches a maximum of approximately $|\Delta P_R|\simeq 0.1$ at $\Delta M\simeq17.4~\mathrm{MeV}$. The benchmark value $\Delta M=16.593~\mathrm{MeV}$ therefore lies close to the resonant maximum and gives
$|\Delta P_R|\simeq0.1$.

\section{BAU calculation with zero-temperature rates and sudden PBH evaporation}
\label{sec:BAU_T0}
To better connect with the results in Ref.~\cite{Ambrosone:2021lsx}, we will follow their assumptions in ignoring the finite-temperature corrections and in assuming that PBHs do not evolve, but instead evaporate instantaneously at temperature $T_{e}$ given by~\cite{Carr:2009jm}
\begin{equation}\label{eq:Te}
T_{e} \simeq 10^{10}~\text{GeV} (M_{\mathrm{g}})^{-3/2}\,,
\end{equation} 
where $M_{\mathrm{g}}$ is the mass of the PBH in grams. Further, we simplify by assuming that the bath energy is $E_{a} = T$, such that COM energy is given by (assuming $a$ is massless)
\begin{equation}\label{eq:repE}
s \simeq m_{X}^{2} + 2m_{X}T\,.
\end{equation}
Thus, we can express the rare production fraction as
\begin{align}\label{eq:B1}
B_{\text{prod}}^{(0)}= \frac{\sigma_{P}(s)}{\sigma_{0}}\,,
\end{align}

where $\sigma_0=f^4/m_X^2$ is adopted as the reference cross section for scattering of $X$ with the thermal plasma, with $f$ being the coupling of $X$ to the thermal bath. Therefore, the effective loss fractions are 
\begin{equation}\label{eq:effLoss}
B_{L} = B_{\text{prod}}^{(0)} P_{L}^{X}\,, \hspace{5mm} \overline{B}_{L} = B_{\text{prod}}^{(0)} P_{L}^{\overline{X}}\,,
\end{equation}
and the production threshold (neglecting the $Y_{1}-Y_{2}$ mass splitting) is
\begin{equation}\label{eq:E_threshold}
E_{a} \geq E_{\text{th}} = \frac{(M_{Y}+m_{b})^{2} - m_{X}^{2}}{2m_{X}} = 59.6~\text{GeV}\,,
\end{equation}
for our benchmark point. Thus, with the representative-energy approximation $E_a=T$, the zero-temperature production fraction $B_{\rm prod}^{(0)}$ vanishes for $x\geq \frac{m_X}{E_{\rm th}}\simeq 2.18$.

\subsection{Absorption by PBHs}
Here, we briefly review the argument in Ref.~\cite{Ambrosone:2021lsx}. In~\cite{Dolgov:2020kqj, Dolgov:2021tsa}, the infall rate of $X$ particles per unit time 	is estimated to be
\begin{equation}\label{eq:infallingRate}
\frac{dN_{X}}{dt} = \frac{4\pi M_{\rm PBH} m_{X} n_{X}}{n_{0}g\sigma_{X}TM_{P}^{2}}\,,
\end{equation}
where $M_{P}$ is the Planck mass, $ g\simeq 100$ is the number of effectively massless degrees of freedom, $\sigma_X$ is the cross section for scattering of $X$ particles with the SM plasma, $n_{0}$ is the number density of each particle in the plasma, $T$ is its temperature, and $n_{X}$ is the number density of particle $X$. An equivalent expression exists for $\overline{X}$, and the cross sections $\sigma_{X}$ and $\sigma_{\overline{X}}$ are slightly different due to CP violation. In Ref.~\cite{Ambrosone:2021lsx}, the difference is parametrized by what they call the CP-violation efficiency
\begin{equation}
\sigma_{X} -  \sigma_{\overline{X}} \simeq \varepsilon\Big(\frac{m_{X}}{T}\Big)f^{2}\sigma_{X}\,.
\end{equation}
In order to better adapt their treatment to our proposed model, we find it more convenient to parametrize the CP-violation in terms of the related CP-odd drag source instead
\begin{equation}\label{eq:drag}
\delta_{\text{drag}}^{(0)} = B_{\text{prod}}^{(0)}\Big(P_{R}^{X} - P_{R}^{\overline{X}}\Big)\,,
\end{equation}
where the $0$ superscript indicates that this is a zero-temperature approximation. Thus, the absorption rates of $X$ and $\overline{X}$ can be expressed as

\begin{align}\label{eq:absorption}
\Big(\frac{dn_{X}}{dt}\Big)_{\text{abs}} & = - \Gamma_{\text{abs}}^{(0)}(T)\Bigg(1-\frac{\delta_{\text{drag}}^{(0)}}{2} \Bigg) n_{X}\,,\\
\Big(\frac{dn_{\overline{X}}}{dt}\Big)_{\text{abs}} & = - \Gamma_{\text{abs}}^{(0)}(T)\Bigg(1+\frac{\delta_{\text{drag}}^{(0)}}{2} \Bigg) n_{\overline{X}}\,,
\end{align} 
where we have defined\footnote{The dependence of $\Gamma_{\rm abs}^{(0)}$ on the cosmological temperatures $T$ and $T_{\rm form}$ should not be confused with finite-temperature corrections. The superscript $(0)$ indicates that these thermal corrections are neglected.}
\begin{equation}
 \Gamma_{\text{abs}}^{(0)}(T) \equiv \frac{4\pi m_{X}^{3} \beta T_{\text{form}}}{g f^{4}T M_{P}^{2}}\,,
\end{equation}
where $\beta = \rho_{\text{PBH}}(T_{\text{form}})/\rho_{\gamma}(T_{\text{form}})$ is the ratio of the PBH density to radiation density at the PBH formation temperature $T_{\text{form}} \simeq 3\times 10^{15}~\text{GeV}(M_{\mathrm{g}})^{-1/2}$~\cite{Masina:2020xhk}. $X$ and $\overline{X}$ maintain their thermal equilibrium with the plasma through the interaction $X + \overline{X} \leftrightarrow \text{SM}$ until $X$ freezes out
\begin{equation}\label{eq:FO}
\Big(\frac{dn_{X}}{dt} \Big)_{\text{int}} = - \langle \sigma_{\rm ann} v \rangle (n_{X} n_{\overline{X}} - n_{\text{eq}}^{2})\,,
\end{equation}
where $ \langle \sigma_{\rm ann} v \rangle \simeq \frac{f^{4}}{m_{X}^{2}}$, and $n_{\rm eq}$ is the equilibrium number density, with an analogous equation for $\overline X$. The order of magnitude of the decay width of $X$ can be expressed as $\Gamma_{X} = h^{2} m_{X}$, with $h$ being the coupling of the  $X$ decay process
\begin{equation}\label{eq:Xdecay}
\Big( \frac{d n_{X}}{dt}\Big)_{\text{decay}} = -h^{2}m_{X} n_{X}\,.
\end{equation}
Again, a similar expression holds for $\overline{X}$. 
\subsection{Transport equations}
Similar to Ref.~\cite{Ambrosone:2021lsx}, we define $N \equiv (n_{X} + n_{\overline{X}})/2s$, $A \equiv (n_{X} - n_{\overline{X}})/s$, and $x \equiv m_{X}/T$. Thus, the evolution equations read

\begin{align}
\frac{dN}{dx} & =
-\frac{\lambda}{x^2} \left(N^2-Y_{\rm eq}^2\right) -\Big[\alpha(x)x^2 +\mu x\Big]N -\frac{r_0(x)}{2}
\left[ B_L\left(N+\frac{A}{2}\right) + \overline{B}_L\left(N-\frac{A}{2}\right) \right]\,,\label{eq:transport1}\\
\frac{dA}{dx} & = \alpha(x)x^2 \left[ \delta_{\rm drag}^{(0)}N-A \right]
-\mu xA -r_0(x) \left[B_L\left(N+\frac{A}{2}\right) -\overline{B}_L\left(N-\frac{A}{2}\right) \right]\,, \label{eq:transport2}
\end{align}
where 
\begin{equation}
\begin{aligned}
&\lambda = \sqrt{\frac{\pi}{45}}\, \frac{f^{4}\sqrt{g_{*}}M_{\rm Pl}}{m_{X}}\,, \quad & \mu & = \sqrt{\frac{45}{4\pi^{3}}}\, \frac{h^{2}M_{\rm Pl}} {\sqrt{g_{*}}m_{X}}\,,\\
&\alpha(x) =\frac{\pi^{7/2}}{\sqrt{5}\,\zeta(3)}\, \frac{\beta T_{\rm form}}{g f^{4}\sqrt{g_{*}}M_{\rm P}}\Theta(x_{e} - x)\,,\quad & r_0(x)& = \frac{n_0(T)\sigma_0}{H(T)x}\,.
\end{aligned}
\end{equation}

Here we work to leading order in the small asymmetry and CP-odd source,
neglecting terms of $\mathcal{O}(A^2)$ and $\mathcal{O}(\delta_{\rm drag}A)$. Notice here that we have introduced the Heaviside step function to model the assumption of an instantaneous PBH evaporation at $T_{e}$, as assumed in Ref.~\cite{Ambrosone:2021lsx}. The baryon asymmetry remaining in the $X$ sector is subsequently transferred to light SM states through $X$ decay, while the loss channel transfers the compensating baryon charge to the spectator sector $F_r$. Thus, we have
\begin{align}
& \frac{d\eta_B^{X\,\mathrm{decay}}}{dx} =
\mu x A\,,\label{eq:BAUeq1} \\[2mm]
& \frac{d\eta_B^{\mathrm{spect}}}{dx} = r_0(x)
\left[ B_L\left(N+\frac{A}{2}\right) -\overline{B}_L\left(N-\frac{A}{2}\right)\right]\,,\label{eq:BAUeq2}
\end{align}
and the total BAU is $\eta_B^{\mathrm{total}} = \eta_B^{X\,\mathrm{decay}}+\eta_B^{\mathrm{spect}}$.

\subsection{Zero-temperature results}
\begin{figure}[t!] 
\centering
\includegraphics[width=0.8\textwidth]{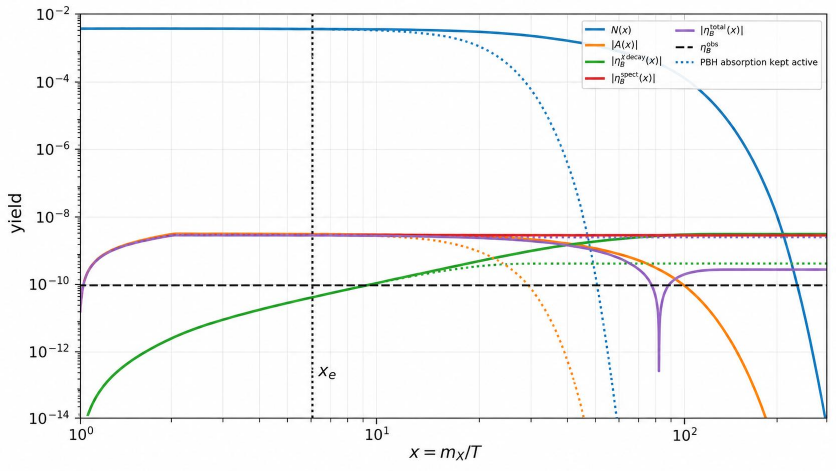}
\caption{Zero-temperature solution of the transport equations. The plot shows the evolution of the $N$ (solid blue), $A$ (solid orange), $\eta_{B}$ from the decay of $X$ (solid green), from the spectator sector (solid red), and the total asymmetry $\eta_B^{\rm total}$ (solid magenta). The dotted lines show the evolution of the respective quantities without PBH decay. The dotted red curve nearly overlaps the corresponding solid red curve, while the dotted magenta is right under the solid red line. The horizontal dashed line denotes the observed BAU and the vertical dotted line indicates the point of PBH evaporation.}
\label{fig:ZeroTsol}
\end{figure}

Here we show the results of the BAU for the benchmark point in Eq.~(\ref{eq:benchmark}). We set $M_{\text{PBH}} = 6 \times 10^{5}~\text{g}$, $\beta = 10^{-12}$, $f = 7 \times 10^{-5}$, $h=3.29 \times 10^{-10}$, and $g_{b} = 1.281 \times 10^{-5}$. Under the instantaneous PBH decay assumption, Eq.~(\ref{eq:Te}) yields $T_{e}^{\text{inst.}} \simeq 21.52~\text{GeV}$, and $x_{e} \simeq 6.04$. The full solution of the transport equations~(\ref{eq:transport1})-(\ref{eq:BAUeq2}) is shown in Figure~\ref{fig:ZeroTsol}. We can see from the plot that while $N$ and $A$ are eventually depleted, the baryon asymmetry nonetheless survives since it is transferred to the SM via the decay of $X$ to light baryonic states and the spectator sector $F_{r}$. Notice that $\eta_B^{X\,\mathrm{decay}}>0$ whereas $\eta_B^{\mathrm{spect}}<0$. Their sum, $\eta_B^{\mathrm{total}}$, is shown by the solid magenta line\footnote{We assume that the baryon charge stored in the spectator sector is subsequently transferred to visible baryonic states, so that $\eta_B^{\rm total}$ corresponds to the late-time observable BAU.}. The dotted lines show the respective quantities had PBHs not evaporated.

Notice that unlike the minimal realization of Ref.~\cite{Ambrosone:2021lsx}, and for our specific benchmark point, the surviving baryon asymmetry if PBH absorption is kept active is actually \textit{larger} than the case where PBHs evaporate! The reason is that if PBH absorption is kept active, then there will be two competing terms that remain: $-\Gamma_{\rm abs}A$ and $+\Gamma_{\rm abs}\delta_{\rm drag}^{(0)}N$, and for this specific benchmark point, the continued source eventually dominates the evolution and drives the total asymmetry through zero and to a large opposite-sign value.

Numerically, the raw produced BAU is $\eta_B^{\text{raw}}\simeq 2.05\times10^{-10}$. The sudden decay of $X$ will lead to entropy injection that dilutes the raw BAU by the factor (see Appendix~\ref{app:C} for the details of entropy dilution assuming a sudden decay)
\begin{equation}\label{eq:EntropyDilution}
D_{s} \equiv \frac{s_{\text{after}}}{s_{\text{before}}} = \Big(1+\frac{\rho_{X}}{\rho_{R}}\Big)^{3/4} = \Big(1+\frac{8}{3}xN\Big)^{3/4}\,, 
\end{equation}
which yields a factor of $D_{s} \simeq 1.174$ at the decay of $X$ around $x\sim 41$. This brings the final BAU to
\begin{equation}
\eta_B^{T=0}
=\frac{\eta_B^{\rm raw}}{D_s}
\simeq1.74\times10^{-10}
\simeq2.00\,\eta_B^{\rm obs}.
\label{eq:zeroT-result}
\end{equation}
which actually \textit{exceeds} the observed BAU! We have intentionally chosen benchmark values to yield this larger value because as we will show in the next sections, when we take into consideration the thermal effects, continuous evaporation of PBHs rather than assuming an abrupt decay, and when accounting for continuous entropy production, the same benchmark values lead to the observed BAU.

\section{Thermal corrections and continuous evolution of PBHs}\label{sec:thermal}
Our earlier treatment made a few simplifying assumptions, the most important of which are that we neglected thermal effects from the plasma and assumed that PBHs evaporate abruptly at $T_e$. Both assumptions need to be revisited. In this section, we incorporate finite-temperature corrections and treat the continuous evaporation of PBHs.
\subsection{Thermal corrections}
\begin{figure}[t!] 
\centering
\includegraphics[width=\textwidth]{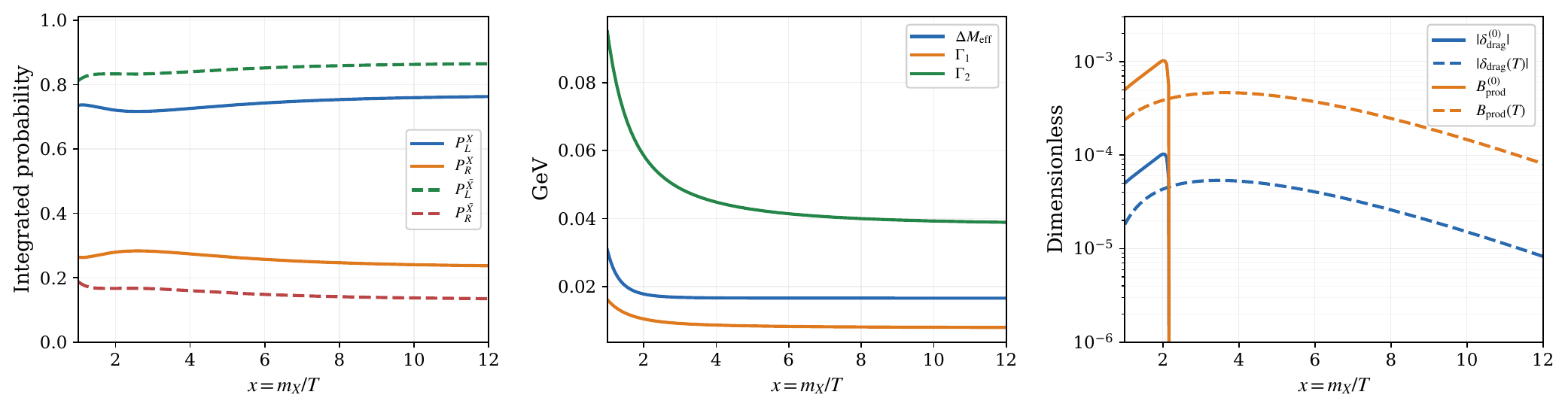}
\caption{Impact of thermal corrections on the integrated return/loss probabilities (left), the effective mass splitting $\Delta M_{\rm eff}$ and widths $\Gamma_{1,2}$ (middle), and the production fraction and CP-odd drag source (right). In the right panel, solid curves show the zero-temperature representative-energy approximation, while dashed curves show the finite-temperature result.}\label{fig:ThermalCorrections}
\end{figure}

Our earlier treatment set $E_{a} = T$, which only serves as a rough estimate. A more careful treatment requires thermal averaging of the incoming bath particles. For a relativistic species $a$, we have
\begin{equation}\label{eq:ThermalB}
B_{\text{prod}}(T) = \frac{2}{3\sigma_{0}\zeta(3)}\int_{0}^{\infty}dy\frac{y^{2}}{1+e^{y}}\sigma_{P}\Big( m_{X}^{2} + 2m_{X} y T\Big)\,,
\end{equation}
where $y = E_{a}/T$. In addition, for the loss and return channels, we need to account for the Bose-Einstein enhancement and Fermi-Dirac blocking. Therefore, when $F_{r}$ and $S_{r}$ thermalize, the thermally-corrected decay widths are given by~\cite{Weldon:1983jn}
\begin{align}
& \Gamma_{L}(T) = \Big( 1- f_{F}(M_{Y}/2,T) + f_{B}(M_{Y}/2,T) \Big)\times \Gamma_{L}(T=0)\,,\\
& \Gamma_{R}(T) = \Big( 1-f_F(E_X,T)+f_B(E_\psi,T)\Big)\times \Gamma_{R}(T=0)\,,\label{eq:gammaR}
\end{align}
where:
\begin{align}
E_{X} & = \frac{M_{Y}^{2} +m_{X}^{2} - m_{\psi}^{2}}{2M_{Y}} = 134.63~\text{GeV}\,,\\
E_{\psi} & = \frac{M_{Y}^{2} - m_{X}^{2} + m_{\psi}^{2}}{2M_{Y}} = 35.37~\text{GeV}\,,
\end{align}
for our benchmark point. Here we assume a thermal distribution for $\psi$ with negligible chemical potential, consistent with the rapid CP-even equilibration discussed in Appendix~\ref{app:G}. Another important point to consider is the thermal corrections to the mass $M_Y$, which is given by
\begin{equation}\label{eq:ThermalMass}
\delta M_{ij}(T) = \frac{T^{2}}{32M_Y} \left[ \sum_{r=1}^{3}\kappa_{ir}\kappa_{jr}^{*}+y_i y_j^{*}\right]\,,
\end{equation}
and we have $M_{\rm eff}=M+\delta M(T)$. Finally, we also have to evaluate the probabilities $P_{c}$ at finite temperature by solving the Sylvester equation and Eq.~(\ref{eq:Pc}). Thus, the CP-source from the return channel becomes
\begin{equation}\label{eq:thermalDrag}
\delta_{\text{drag}}(T) = B_{\text{prod}}(T)[P_{R}^{X}(T) - P_{R}^{\overline{X}}(T)]\,.
\end{equation}
Notice here that the compensating loss difference becomes $\delta_{\text{L}}(T) = B_{\text{prod}}(T)[P_{L}^{X}(T) - P_{L}^{\overline{X}}(T)]$, and that by probability conservation and CPT invariance one should have $\delta_{\text{drag}}(T)  + \delta_{\text{L}}(T) =0$. We have checked numerically that the cancellation holds to an accuracy of $\mathcal{O}(10^{-12})$. We show the impact of the thermal corrections in Figure~\ref{fig:ThermalCorrections}.

\subsection{Continuous PBH evaporation}
The previous treatment assumed that PBHs evaporate abruptly at $T_{e}$ given by Eq.~(\ref{eq:Te}), which is not a very realistic assumption. Instead, we account for continuous PBH evaporation using the hot-PBH mass-loss approximation~\cite{MacGibbon:1991tj}
\begin{equation}
\frac{dM_{\mathrm{g}}}{dt} = -5.34 \times 10^{25} F(M) \Big( \frac{ M_{\mathrm{g}}}{ \mathrm{g}}\Big)^{-2} ~\mathrm{g} s^{-1}\,,
\end{equation}
where $F(M_{\mathrm{g}}) \simeq 15.35$\footnote{For simplicity, we use the Standard-Model high-temperature value $F(M_{\mathrm{g}})\simeq 15.35$ and neglect the additional contribution of the new degrees of freedom introduced in our model. For the field content considered here, this contribution is estimated to be
$\Delta F_{\rm BSM}\sim 5$, corresponding to an $\mathcal{O}(30\%)$ correction to the Hawking mass-loss rate.} is the dimensionless evaporation function encoding the particle species accessible to Hawking emission and their spin-dependent greybody factors, and $\dot n_{\rm PBH}+3Hn_{\rm PBH}=0$. Thus, the energy lost per unit volume is injected into radiation
\begin{equation}
Q_{\text{PBH}} = -c^{2} \dot{M}_{\mathrm{g}}n_{\text{PBH}}\,.
\end{equation}

\subsection{Updated transport equations and results}
\begin{figure}[t!] 
\centering
\includegraphics[width=\textwidth]{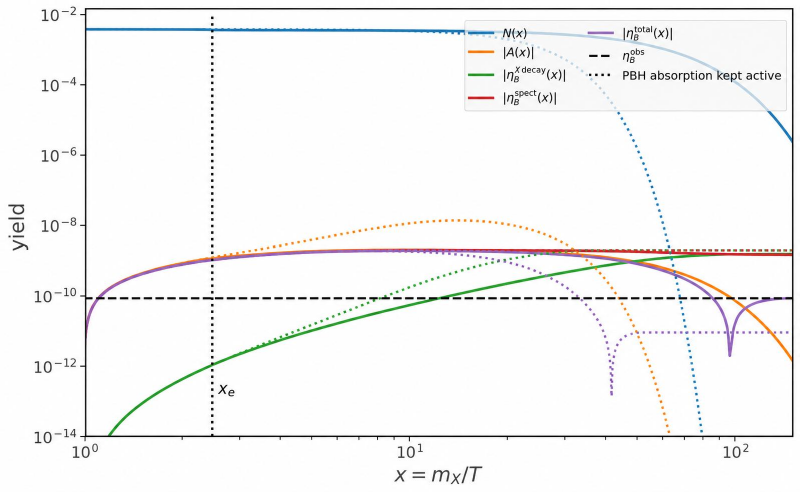}
\caption{Solution of the updated transport equations. Solid curves show the physical evolution with continuous PBH evaporation, while dotted curves show the corresponding evolution when PBH absorption is
artificially kept active. The colors denote $N$, $A$, $\eta_B^{X\,\mathrm{decay}}$, $\eta_B^{\mathrm{spect}}$, and $\eta_B^{\mathrm{total}}$ as indicated in the legend.}\label{fig:Tsol}
\end{figure}

As entropy is no longer assumed to be constant, it is more convenient to rewrite the transport system in  terms of $n_{\pm}$ defined as

\begin{equation}\label{eq:npm}
n_{+} = \frac{n_{X}+n_{\overline{X}}}{2}\,, \hspace{5mm} n_{-} = n_{X}-n_{\overline{X}}\,.
\end{equation}
Thus, the corrected transport equations become

\begin{align}
& \dot n_+ =
-3Hn_+ -\langle \sigma_0 v\rangle \left( n_Xn_{\bar X}-n_{\rm eq}^2 \right) -\Gamma_{\rm abs}n_+ -\Gamma_X n_+ -\frac{1}{2} \left( \Gamma_L^X n_X + \Gamma_L^{\bar X} n_{\bar X} \right)\,,\\
& \dot n_- = -3Hn_- +\Gamma_{\rm abs}\delta_{\rm drag}n_+ -\Gamma_{\rm abs}n_- -\Gamma_X n_- -\left( \Gamma_L^X n_X - \Gamma_L^{\bar X} n_{\bar X} \right)\,,\\
& \dot n_B = -3Hn_B + \Gamma_X n_-\,,\\
& \dot q_B = -3Hq_B + \left( \Gamma_L^X n_X - \Gamma_L^{\bar X} n_{\bar X} \right)\,,
\end{align}
where $q_{B} = \sum_{r}B(F_{r})(n_{F_{r}} - n_{\overline{F}_{r}})$ is the density of the baryon charge stored in the spectator sector, and the loss rates are given by
\begin{equation}
\Gamma_L^X = n_0\sigma_0 B_{\rm prod} P_L^X, \qquad \Gamma_L^{\bar X} = n_0\sigma_0 B_{\rm prod} P_L^{\bar X}\,.
\end{equation}
Finally, the radiation evolves as
\begin{equation}
\dot\rho_R + 4H\rho_R = Q_{\rm PBH} + 2m_X\Gamma_X n_+\,,
\end{equation}
and the entropy is computed dynamically from $S=sa^3$. We continue to work to leading order in the CP-odd quantities and neglect the term $\frac{1}{4}\Gamma_{\rm abs}\delta_{\rm drag}n_-$. The solution is shown in Figure~\ref{fig:Tsol}, from which we can see that after taking into consideration the above corrections, the resulting BAU matches the experimentally observed value
\begin{equation}
\eta_B^{\rm full}\equiv\frac{n_B+q_B}{s}= 8.72 \times 10^{-11} = \eta_{B}^{\text{obs}}\,,
\end{equation}
which is about $50\%$ smaller than the value obtained at $T=0$. Notice that if PBHs remain active and do not evaporate, we would still have
\begin{equation}
\eta_B^{\rm no\,evap}
\simeq 9.1\times10^{-12}
\simeq 0.105\,\eta_B^{\rm obs}\,.
\end{equation}
The contrast with the zero-temperature control arises because thermal averaging, temperature-dependent coherent probabilities, and continuous PBH and entropy evolution change the relative time dependence of the CP-odd source and depletion terms, so that continued PBH absorption suppresses rather than enhances the late-time total asymmetry in the full treatment. 

For comparison, we find here that the entropy dilution factor is $D_{s} \simeq 1.43$, which, although significantly larger than the previous value of $\sim 1.174$, remains modest.
To verify that our solution is under control, we first compare the PBH energy density with the radiation energy density. We find that $\max{(\rho_{\text{PBH}}/\rho_{R})} \simeq 4.21 \times 10^{-2} \ll 1$, which implies radiation domination throughout the full evolution. We also find that $T_{e} \simeq 52.6~\text{GeV} \gg T_{\text{BBN}}$ and thus BBN is unaffected.  In addition, we find $\Gamma_{\rm prod}/H\sim10^{-6}$, indicating that the rare $\phi$-mediated production process remains out of equilibrium. This does not imply that the complete reaction network is out of equilibrium, as discussed in Appendix~\ref{app:G}. Finally, we compare the PBH accretion rate with the Hawking evaporation rate. We find that $\max{R_{ae}} \simeq 7.9 \times 10^{-4} \ll 0.1$, which implies that PBH accretion remains negligible relative to their decay.  

We have also investigated the additional inverse reactions and annihilation channels implied by the Lagrangian. As detailed in Appendix~\ref{app:G}, these processes can maintain chemical equilibrium during the PBH-capture epoch. Using the Bose-corrected resonant source and replacing the reference $X$ abundance by its equilibrium value, while retaining the effective capture and transport prescriptions, yields an abundance-only source reduction of approximately $33.5\%$. Thus, the corresponding Boltzmann suppression is at most of $\mathcal{O}(1)$.

\section{Detection with Gravitational Waves}\label{sec:GW}

\begin{figure}[t!] 
\centering
\includegraphics[width=\textwidth]{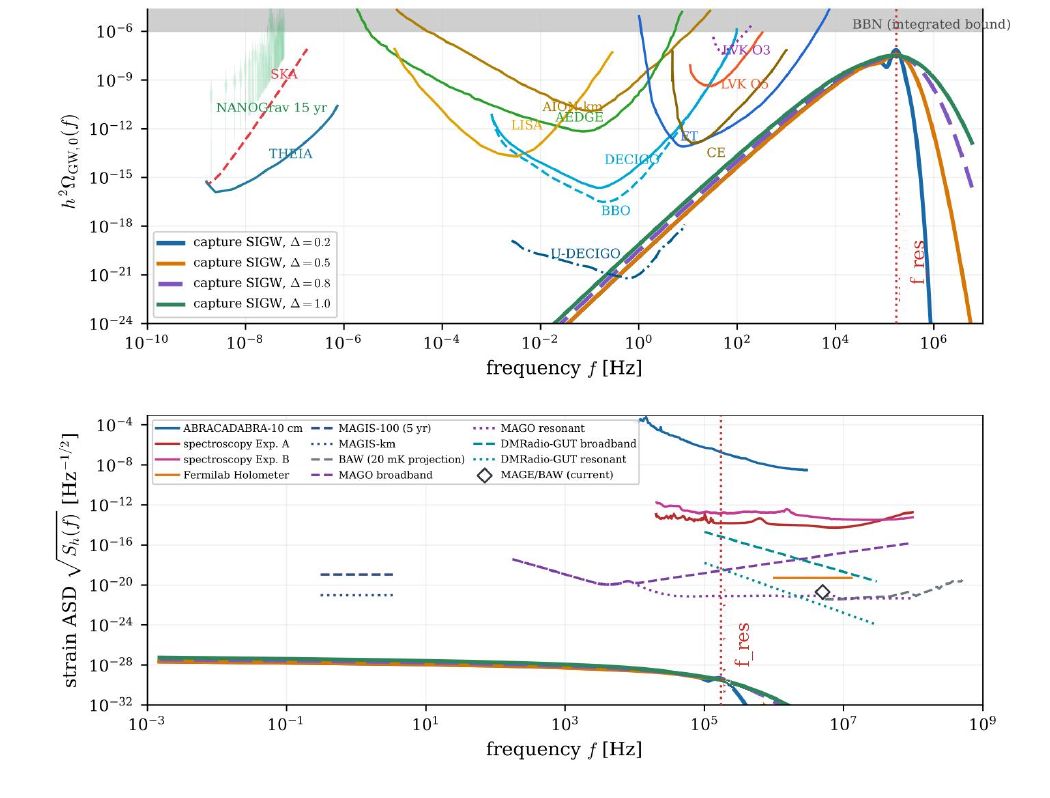}
\caption{The SIGW power spectrum compared to the relevant experiments. (Top): $h^{2}\Omega_{GW}(f)$ with conventional stochastic-reach curves and the BBN integrated bound~\cite{Caprini:2018mtu}. (Bottom): current and projected mid/high-frequency native strain ASDs~\cite{Coleman:2018ozp, Pappas:2025zld, Budker:2026uec, Holometer:2016qoh, Goryachev:2021zzn, Berlin:2023grv,DMRadio:2022jfv, Domcke:2023bat}. The red dotted line marks the central benchmark resonance scale $f_{\rm res}\simeq1.73\times10^{5}~{\rm Hz}$. The diamond represents the current MAGE/BAW (bulk acoustic wave) experimental constraint/sensitivity point.}
\label{fig:GW}
\end{figure}
In this section, we explore the possibility of probing the resonantly-enhanced BAU production via GWs associated with the collapse of the PBHs required to make this mechanism succeed. For PBHs formed by the collapse of an enhanced primordial curvature spectrum, the most direct GW source is the Scalar-Induced GW (SIGW) background generated at second order when those scalar modes re-enter the horizon during radiation domination~\cite{Ananda:2006af, Baumann:2007zm, Kohri:2018awv, Espinosa:2018eve}. Direct Hawking gravitons are unavoidable in principle, but their spectrum is controlled by a different frequency scale and by greybody and spin physics~\cite{Dolgov:2011cq,Ireland:2023avg} and we neglect them here. For the modes re-entering the horizon, we assume that a fraction $\gamma$ of the mass in the horizon collapses into a PBH, such that $M = \gamma M_{H}$. In our calculation, we set $\gamma = 0.2$. Entropy conservation gives the present frequency of the scalar mode that formed the PBH~\cite{Kohri:2024qpd, Espinosa:2018eve}
\begin{equation}\label{eq:f_star}
f_{*} \simeq 3.67 \times 10^{5}~\text{Hz} \Big(\frac{M_{\mathrm{g}}}{10^{5} \mathrm{g}}\Big)^{-1/2} \Big(\frac{\gamma}{0.2}\Big)^{1/2} \Big(\frac{g_{*}}{106.75}\Big)^{-1/12}\,,
\end{equation}
however, for a narrow scalar peak, the radiation-era kernel resonates near $k\simeq 2k_{*}/\sqrt{3}$~\cite{Kohri:2018awv, Espinosa:2018eve}. Therefore
\begin{equation}\label{eq:f_res}
f_{\text{res}} \simeq \frac{2}{\sqrt{3}} f_{*}\,.
\end{equation}
We parametrize the curvature enhancement as 
\begin{equation}\label{eq:curvature_prob}
\mathcal{P}_{\zeta}(k) = A \exp{\Bigg[\frac{-\ln^{2}(k/k_{*})}{2\Delta^{2}} \Bigg]}\,,
\end{equation}
and for Gaussian collapse, the PBH abundance fraction can be modeled as
\begin{equation}
\beta = \frac{\gamma}{2} \text{erfc}\Big(\frac{\delta_{c}}{\sqrt{2}\sigma_{\delta}} \Big)\,,
\end{equation}
\begin{figure}[t!] 
\centering
\includegraphics[width=.8\textwidth]{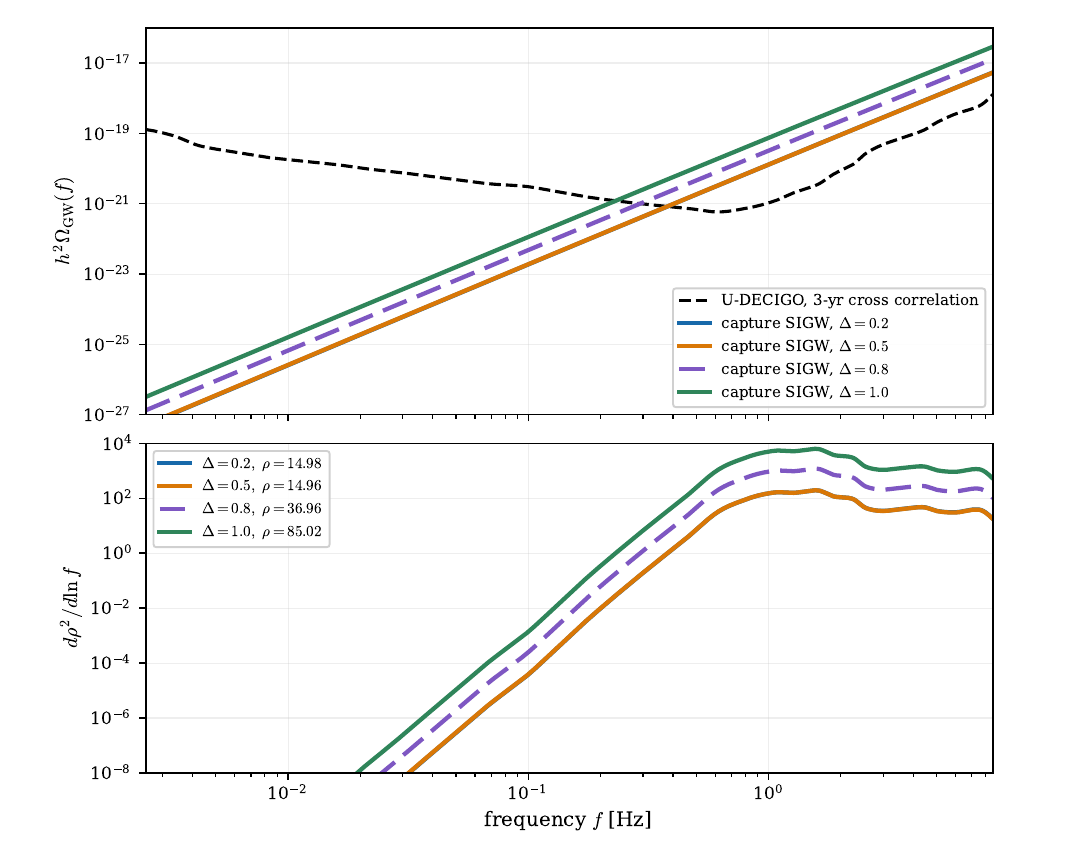}
\caption{(Top): a zoomed-in view of the three-year U-DECIGO characteristic curve and the SIGW power spectrum of the 4 benchmark points. (Bottom): the contribution $\frac{d\rho^{2}}{d\ln{f}}$ vs the frequency. The entries give the integrated three-year SNR. The blue and green lines are on top of each other.}
\label{UDECIGO:GW}
\end{figure}
where $\text{erfc}(x)$ is the complementary error function, $\sigma^{2}_{\delta}$ is the variance of the smoothed radiation-density perturbation, and we set the threshold $\delta_{c} = 0.45$. We show the technical details for the SIGW calculation in Appendix~\ref{app:E}. We calculate the GW power spectrum for a few benchmark points corresponding to $\Delta = (0.2, 0.5, 0.8, 1)$ and show the results in the top panel of Figure~\ref{fig:GW} and superimpose the sensitivity curves of relevant GW experiments. We can see from the plot that typically the amplitudes range between $\sim (3-7) \times 10^{-8}$, and that the resonant peaks lie around $\sim 0.16-0.17$ MHz, which is well above the frequency bands of conventional space- and ground-based interferometers, including the mHz--Hz range commonly associated with
electroweak-scale First-Order Phase Transitions (FOPTs)~\cite{Abu-Ajamieh:2026ndt}. Experiments that are sensitive to the kHz-MHz range typically have their sensitivity curves expressed in terms of the Amplitude Spectral Density (ASD) of gravitational-wave strain, which is related to the GW power spectrum as
\begin{equation}\label{eq:ASD}
\Omega_{\text{GW}}h^{2}(f) =\frac{2\pi^2}{3H_{100}^2}f^3S_h(f)\,,
\end{equation}
where $H_{100} = 100~\text{km} \text{s}^{-1}\text{Mpc}^{-1}$, $S_{h}$ is the one-sided strain power spectral density, and $\text{ASD}(f) = \sqrt{S_{h}(f)}$. We show the strain ASD in the lower panel of Figure~\ref{fig:GW} and superimpose the sensitivity curves of relevant experiments for that frequency range. 

As Figure~\ref{fig:GW} shows, among the experiments considered here, Ultimate DECIGO is the only one
with appreciable sensitivity to the finite-width infrared tail of the SIGW signal. Although the SIGW peak lies near $\sim 0.16$ MHz, its finite-width infrared tail extends into the U-DECIGO band and can be probed under the idealized sensitivity assumptions adopted here. We further investigate its sensitivity by calculating the Signal-to-Noise (SNR) ratio over the 3-year cross-correlation characteristic curve
\begin{equation}
\rho(3~\text{yr}) = \sqrt{\int d\ln{f} \Bigg[\frac{h^{2}\Omega_{\text{SIGW}}(f)}{h^{2}\Omega^{(3\text{yr})}_{\text{UDECIGO}}(f)} \Bigg]^{2}}\,,
\end{equation}
which is the characteristic-curve form of the standard stochastic cross-correlation expression~\cite{Kuroyanagi:2014qza}. Assuming noise-dominated scaling, $\rho(T) = \rho(3\text{yr})\sqrt{T/(3\text{yr})}$. In the top panel of Figure~\ref{UDECIGO:GW}, we show a zoomed-in view of the overlap region between the SIGW power spectrum of the benchmarks and the U-DECIGO sensitivity curve, and we show in the bottom panel the differential SNR vs frequency. The results for the four spectral-width benchmarks are summarized in Table~\ref{tab1}. As the plots and table show, U-DECIGO can potentially detect the SIGW from the candidate PBHs with an SNR exceeding 10 for all benchmark points.

\section{Discussion}
\label{sec:Discussion}

We have shown that the asymmetric-capture mechanism of baryogenesis from primordial black holes can reproduce the observed baryon asymmetry once the small loop-induced CP source of earlier constructions is replaced by a resonantly-enhanced source. The key ingredient is the coherent production and mixing of two nearly degenerate unstable states, $Y_1$ and $Y_2$, followed by CP-asymmetric return and loss decays. For the benchmark considered here, the full finite-temperature treatment with continuous PBH evaporation and entropy production gives
\begin{equation}
\eta_B^{\rm full}
=
8.72\times10^{-11}\,,
\end{equation}
in agreement with the observed BAU. The solution remains under control: the Universe stays radiation dominated, PBH accretion is negligible, the rare production process remains out of equilibrium, and the CPT/probability-conservation closure relation between return and loss probabilities is satisfied numerically to high accuracy.

To verify that our results are not restricted to a single finely selected benchmark, we have extended the numerical analysis to a broader set of microscopic and cosmological parameters, as detailed in Appendix~\ref{app:F}. In particular, we have varied the resonant mass splitting, CP-violating phases, interaction couplings, PBH masses and initial abundances. We find twelve additional benchmark points reproducing the observed BAU within the reference transport treatment. Furthermore, variations of individual parameters without retuning the production coupling yield asymmetries of the same order of magnitude, indicating that successful baryogenesis is not confined to an isolated numerical point within the resonant regime.

\begin{table}[t!]
\centering
\begin{tabular}{cccccc}
\hline
$\Delta$ & $A$ & $f_{\rm peak}\,[\mathrm{kHz}]$ & $h^2\Omega_{\rm peak}$ & $h_c^{\rm peak}$ & $\rho_{3\,\mathrm{yr}}$ \\
\hline
0.2 & 0.1324 & 170.2 & $7.18\times10^{-8}$ & $1.98\times10^{-27}$ & 14.98 \\
0.5 & 0.06804 & 170.2 & $3.70\times10^{-8}$ & $1.42\times10^{-27}$ & 14.96 \\
0.8 & 0.05769 & 161.1 & $3.22\times10^{-8}$ & $1.40\times10^{-27}$ & 36.96 \\
1.0 & 0.05515 & 157.2 & $3.17\times10^{-8}$ & $1.43\times10^{-27}$ & 85.02 \\
\hline
\end{tabular}
\caption{SIGW peak quantities and the projected three-year SNR of U-DECIGO statistic for the benchmark points.}
\label{tab1}
\end{table}

We also explored the gravitational-wave signatures associated with the primordial perturbations responsible for forming the required PBHs. The resulting scalar-induced gravitational-wave spectra peak around
\begin{equation}
f_{\rm peak}\sim0.16\text{--}0.17~{\rm MHz}\,,
\qquad
h^2\Omega_{\rm GW}^{\rm peak}\sim(3\text{--}7)\times10^{-8}\,.
\end{equation}

Although these gravitational waves are not unique to the baryogenesis mechanism itself, they provide an independent probe of the PBH formation history required by the scenario. Under the idealized sensitivity assumptions adopted here, U-DECIGO can achieve SNR $ > 10$ for the benchmark spectra. Overall, resonant enhancement therefore provides a viable way to overcome the suppression that limited previous PBH asymmetric-capture models and opens a complementary gravitational-wave window on the required primordial-black-hole population.

\section*{Acknowledgements}
We thank SOKN Engineering for their hospitality and support. We have used ChatGPT to help write the code, prepare the plots and revise the manuscript. 

\appendix

\section{Appendix A: kinematics}\label{app:A}
For $X\bar a \to Y\bar b$ with a massless $a$, the exact center-of-mass quantities are
\begin{align}
E_i &= \frac{s+m_X^2}{2\sqrt{s}}\,, 
&\qquad p_i &= \frac{s-m_X^2}{2\sqrt{s}}\,, \\
E_f &= \frac{s+M_Y^2-m_b^2}{2\sqrt{s}}\,, 
&\qquad p_f &= \frac{\lambda^{1/2}(s,M_Y^2,m_b^2)}{2\sqrt{s}}\,, \\
t_0 &= m_X^2+M_Y^2-2E_iE_f, 
&\qquad t_{\pm} &= t_0\pm2p_ip_f\,.
\end{align}
Using
\begin{equation}
p\cdot k=\frac{m_X^2+M_Y^2-t}{2}\,, \qquad q\cdot\ell=\frac{m_b^2-t}{2}\,, \qquad  s \geq (M_{Y} + m_{b})^{2}\,.
\end{equation}
and for the corrected $s$-channel $\phi$-mediated production process, the spin-averaged matrix element is independent of $t$, so the two-body cross section can be evaluated analytically. 

\section{Appendix B: derivation of the Sylvester equation}\label{app:B}
Starting from
\begin{equation}
\rho(t)=e^{-iHt}pp^\dagger e^{+iH^\dagger t}\,,
\end{equation}
the integrated probability into channel $c$ is
\begin{equation}
P_c=p^\dagger K_c p, \qquad
K_c\equiv \int_0^\infty dt\, e^{+iH^\dagger t}\Gamma_c e^{-iHt}\,.
\end{equation}
Differentiating the integrand and using the exponential damping of the unstable system at $t\to\infty$ gives
\begin{equation}
H^\dagger K_c-K_cH=i\Gamma_c.
\end{equation}
Summing over $c=L,R$ and noting that $H^\dagger-H=i(\Gamma_L+\Gamma_R)$, we obtain
\begin{equation}
K_L+K_R=\mathbb{1}\,.
\end{equation}

\section{Appendix C: sudden decay entropy dilution}\label{app:C}
Here we calculate the entropy dilution assuming sudden decay of $X$ and $\overline{X}$. Since at the time of decay they are nonrelativistic, their density is given by
\begin{equation}
\rho_{X} \simeq m_{X} (n_{X} + n_{\overline{X}}) = 2m_{X} sN\,,
\end{equation}
given that $s = \frac{2\pi^{2}}{45}g_{*s}T^{3}$ and $\rho_{R} = \frac{\pi^{2}}{30}g_{*}T^{4}$, we have
\begin{equation}
\frac{\rho_{X}}{\rho_{R}} \simeq \frac{8}{3} x N \,,
\end{equation}
where we have assumed that $g_{*s} \simeq g_{*}$. Assuming that $X$ and $\overline{X}$ decay to radiation that quickly thermalize with the plasma, we have
\begin{equation}
\frac{\rho_{\text{after}}}{\rho_{\text{before}}} = \frac{\rho_{R}+\rho_{X}}{\rho_{R}} = \Big(1 + \frac{\rho_{X}}{\rho_{R}} \Big) \implies \frac{T_{\text{after}}}{T_{\text{before}}} = \Big(1 + \frac{\rho_{X}}{\rho_{R}} \Big)^{1/4}\,.
\end{equation}
since $s \propto T^{3}$, and neglecting the change in $g_{*s}$ across the sudden decay, the entropy dilution factor is given by
\begin{equation}
D_{s} \equiv \frac{s_{\text{after}}}{s_{\text{before}}} = \Big(1 + \frac{\rho_{X}}{\rho_{R}} \Big)^{3/4} =  \Big(1 + \frac{8}{3}x N \Big)^{3/4}\,.
\end{equation}
The decay equation is given by
\begin{equation}
\frac{dN}{dx}\Big|_{\text{dec}} = - \mu x N\,,
\end{equation}
which using $d/dt = H xd/dx$ can be written as
\begin{equation}
\frac{\Gamma_{X}}{H} = \mu x^{2}\,.
\end{equation}
Since the decay becomes cosmologically relevant when $\Gamma_{X} \sim H$, we have $\mu x^{2}_{\text{dec}} \sim 1$, which for the benchmark we use gives $x_{\text{dec}} \simeq 41$ and $N(x_{\text{dec}}) \simeq 2.2 \times 10^{-3}$, which leads to the entropy dilution $D_{s} \sim 1.174$ in the sudden-decay approximation.
\section{Appendix D: comparison of the direct and resonant CP-odd sources}
\label{app:D}
\begin{figure}[t!]
    \centering
    \includegraphics[width=0.78\textwidth]{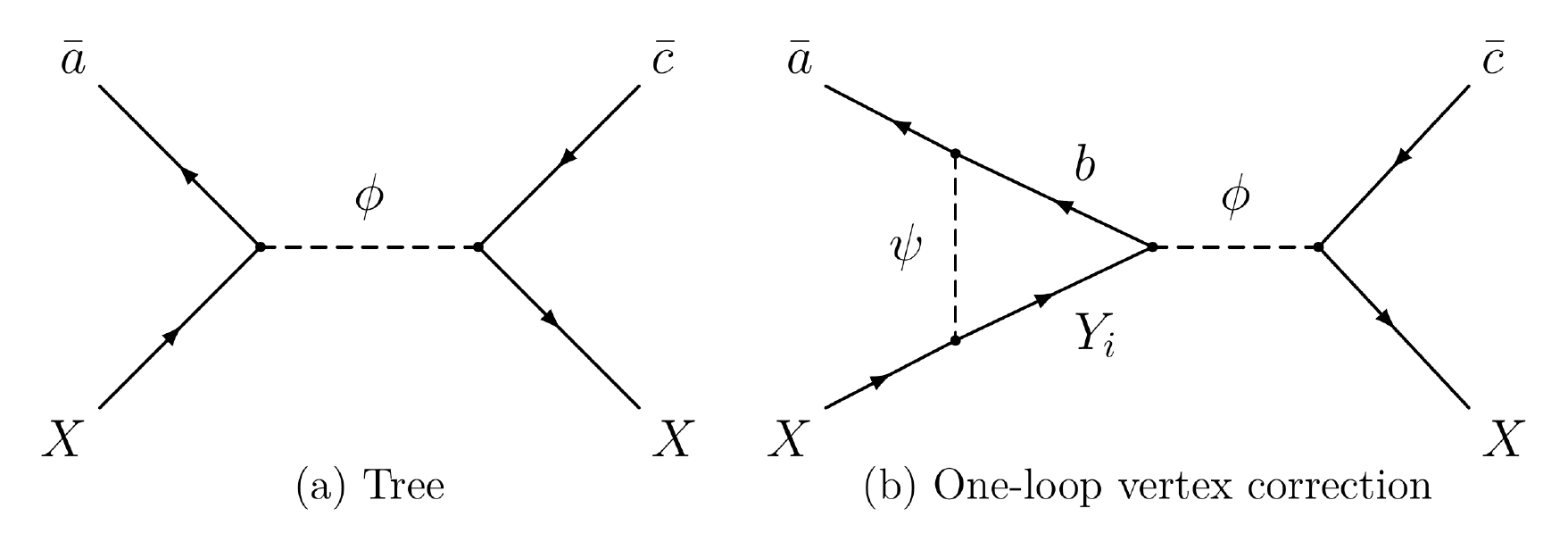}
    \caption{Feynman diagrams that contribute to the scattering $X\overline{a} \rightarrow X\overline{c}$, which represents the main contribution to the BAU in Ref.~\cite{Ambrosone:2021lsx}}
    \label{fig:OriginalProcesses}
\end{figure}
Here we compare the resonantly-enhanced mechanism shown in Figure~\ref{fig:feynman} with the nonresonant CP-odd contribution arising from the interference between the tree-level and one-loop processes shown in Figure~\ref{fig:OriginalProcesses}, following the construction of Ref.~\cite{Ambrosone:2021lsx}. For each intermediate state $Y_i$, the CP-odd tree--loop contribution can be written as
\begin{equation}
\frac{\Delta\sigma_{\rm vertex}(s)}{\sigma_{\rm tree}(s)} = \frac{1}{|g_{aX}|^2} \sum_{i=1}^{2} {\rm Im}
\left(g_{aX}^{*}y_i g_{bY_i}g_{ba}^{*} \right) \mathcal{F}(s,M_{Y_i},m_\psi) \Theta\!\left[s-(m_b+M_{Y_i})^2\right]\,.
\label{eq:Dvertex}
\end{equation}
\begin{figure}[t!]
    \centering
    \includegraphics[width=0.78\textwidth]{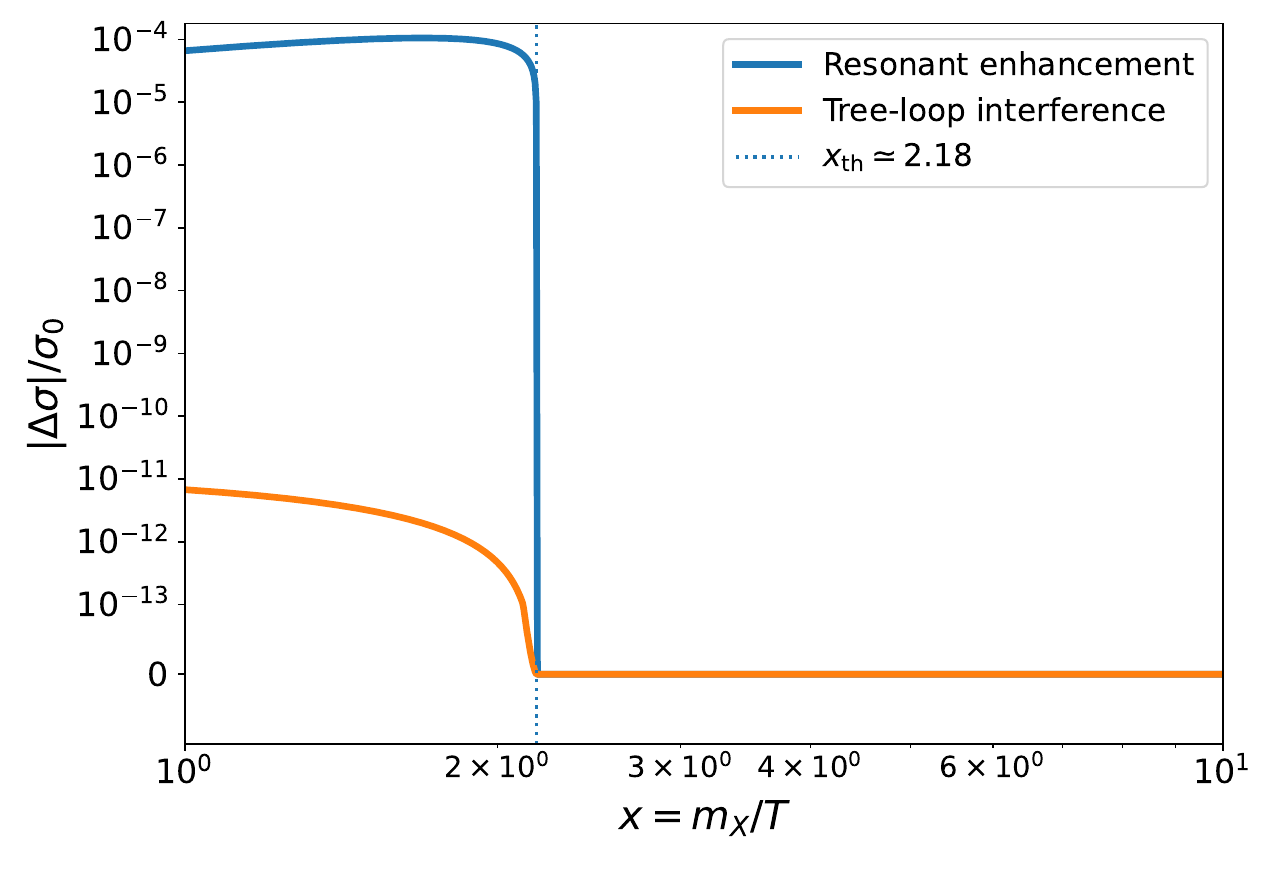}
    \caption{Comparison of the direct vertex CP-odd source with the resonant return/loss source at zero 	temperature, normalized to $\sigma_0=f^4/m_X^2$.}
    \label{fig:comparison}
\end{figure}
For the light-$\psi$ spectrum used in this work, the full propagator must be retained. Cutting the $b$ and $Y_i$ lines gives
\begin{equation}
\mathcal{F}(s,M_i,m_\psi) = -\frac{1}{\pi(s-m_X^2)^2} \int_{t_-}^{t_+}dt\, \frac{\mathcal{N}_i(s,t)} {t-m_\psi^2}\,,
\label{eq:DfullF}
\end{equation}
where
\begin{align}
\mathcal{N}_i(s,t) = \frac{1}{2} \Big\{& m_b \Big[ M_i\big(m_X(m_b+m_X)-s\big) + m_X\big(m_b^2+m_bm_X+m_X^2-s\big) \Big] \nonumber\\
&- \big[s+m_X(M_i+m_b)\big]t \Big\}\,.
\label{eq:DN}
\end{align}
The limits $t_\pm$ are those given in Appendix~\ref{app:A}. For our spectrum, the other possible two-particle cuts are kinematically closed:
\begin{equation}
m_a^2 < (m_b+m_\psi)^2, \qquad m_X^2 < (M_i+m_\psi)^2\,,
\end{equation}
so that the $bY_i$ cut is the only open normal threshold relevant to the absorptive part. As a consistency check, in the heavy-$\psi$ limit $m_\psi^2\gg |t_\pm|$, Eq.~(\ref{eq:DfullF}) reduces to
\begin{equation}
\mathcal{F}(s,M_i,m_\psi) \longrightarrow \frac{ [s-(m_b-M_i)m_X][s-(m_b+M_i)^2]}{4\pi s^2m_\psi^2}\sqrt{
[s-(m_b+M_i)^2][s-(m_b-M_i)^2]}\,,
\end{equation}
which reproduces the result of Ref.~\cite{Ambrosone:2021lsx}. The resonant contribution is instead
\begin{equation}
\Delta\sigma_{\rm res}(s) =\sigma_P(s) \left(P_R^X-P_R^{\overline X} \right)\,.
\label{eq:Dres}
\end{equation}

For the comparison in Figure~\ref{fig:comparison}, we use the benchmark masses and phases of Eq.~(\ref{eq:benchmark}), take the additional direct-channel couplings to be real with $g_{bY_i}/g_{aX}\sim 0.21$, and normalize the tree contribution to $\sigma_{\rm tree}=\sigma_0$. With these choices, $y_1$ gives no CP-odd contribution and the direct source is controlled by the phase of $y_2$. Both contributions vanish once the $bY_i$ production threshold closes, at
\begin{equation}
x_{\rm th} = \frac{m_X}{E_{\rm th}} \simeq 2.18\,.
\end{equation}
Figure~\ref{fig:comparison} shows that the light-$\psi$ tree--loop contribution is strongly suppressed relative to the resonant return/loss source, where we have set $g_{ba} = 2.68 \times 10^{-9}$. For example, at $x=1$ we obtain
\begin{equation}
\frac{|\Delta\sigma_{\rm vertex}|}{\sigma_0} \simeq 1.44\times10^{-12}, \qquad \frac{|\Delta\sigma_{\rm res}|}{\sigma_0} \simeq 6.87\times10^{-5}\,
\end{equation}
so that the resonant contribution is larger by approximately eight orders of magnitude.

\section{Appendix E: technical details of the SIGW calculation}\label{app:E}
The curvature-spectrum inversion uses a Gaussian window $W(z)=e^{-z^2/2}$, and the radiation transfer function is given by~\cite{Ando:2018qdb}
\begin{equation}
T(z) = 3\Big(\frac{\sin z-z\cos z}{z^3}\Big), \qquad R=k_*^{-1}\,,
\end{equation}
where the smoothing scale is chosen to correspond to the central PBH-forming mode. Therefore,
\begin{equation}
\sigma_\delta^2 = \frac{16}{81}\int d\ln q\, (qR)^4 W^2(qR) T^2 \left( \frac{qR}{\sqrt{3}} \right) \mathcal P_\zeta(q)\,.
\end{equation}
The induced tensor density in radiation domination is evaluated as~\cite{Kohri:2018awv, Espinosa:2018eve}
\begin{equation}
\Omega_{\rm GW,RD}(k) =\frac{1}{6} \int_0^\infty dv \int_{|1-v|}^{1+v}du \left[\frac{4v^2-(1+v^2-u^2)^2}{4uv}\right]^2 \overline{I^2(u,v)} \mathcal P_\zeta(ku)\mathcal P_\zeta(kv)\,,
\end{equation}
where using $D=u^2+v^2-3$,
\begin{equation}
\overline{I^2} = \frac{1}{2} \left( \frac{3D}{4u^3v^3} \right)^2 \left\{ \left[-4uv+ D \ln\left|\frac{3-(u+v)^2}{3-(u-v)^2} \right| \right]^2+\pi^2D^2\Theta(u+v-\sqrt{3})\right\}\,.
\end{equation}
The present density is
\begin{equation}
h^2\Omega_{\rm GW,0}(k)=0.83\left(\frac{g_*}{10.75}\right)^{-1/3}h^2\Omega_{r,0} \Omega_{\rm GW,RD}(k)\,,
\qquad h^2\Omega_{r,0} = 4.18\times10^{-5}\,.
\end{equation}

For stable infrared integration we use $ t=u+v$, $s=u-v$, and $y=\left(\frac{k}{k_*}\right)t$, with explicit refinement around the logarithmic resonance $ t=\sqrt{3}$. This keeps the support near $y\simeq 2$ even when $ k\ll k_*$, and avoids replacing the infrared tail by a hand-drawn extrapolation. The finite-width deep infrared behavior approaches $\Omega_{\rm GW} \propto f^3 \ln^2 \left( \frac{f_c}{f} \right)$~\cite{Yuan:2019wwo}.

\section{Appendix F: Additional baryogenesis benchmarks and GW mapping}
\label{app:F}

To investigate the robustness of our results, we first vary the initial
PBH mass while keeping $\beta=10^{-12}$ and all microscopic parameters
fixed, except for the rare-production coupling $g_b\equiv g_{bY_1}=g_{bY_2}$,
which is retuned to reproduce the observed baryon asymmetry. As in
Section~\ref{sec:thermal}, the reference transport calculation includes a thermal
$\psi$ population in the resonant return damping width. The resulting
four benchmark points are given in Table~\ref{tab:appF-original}.

\begin{table}[t!]
\centering\small
\resizebox{\textwidth}{!}{%
\begin{tabular}{ccccccc}
\hline\hline
$M_{\rm PBH}$ [g] & $g_b$ & $\eta_B$ & $T_{\rm evap}$ [GeV] &
$\max(\rho_{\rm PBH}/\rho_R)$ & $f_{\rm res}$ [kHz] & $S_f/S_i$\\
\hline
$4.5\times10^5$ & $3.11962\times10^{-5}$ & $8.720\times10^{-11}$ & 81.05 & 0.0316 & 199.9 & 1.414\\
$6.0\times10^5$ & $1.28126\times10^{-5}$ & $8.720\times10^{-11}$ & 52.55 & 0.0421 & 173.1 & 1.427\\
$8.0\times10^5$ & $6.30048\times10^{-6}$ & $8.720\times10^{-11}$ & 34.09 & 0.0563 & 149.9 & 1.442\\
$1.0\times10^6$ & $3.98898\times10^{-6}$ & $8.720\times10^{-11}$ & 24.37 & 0.0704 & 134.1 & 1.457\\
\hline\hline
\end{tabular}}
\caption{PBH-mass benchmarks with a thermal $\psi$ population included
in the return damping width. All other microscopic parameters are fixed
and only $g_b$ is retuned. The PBH-formation frequencies and idealized
GW sensitivity estimates are unchanged for fixed $M_{\rm PBH}$ and $\beta$.}
\label{tab:appF-original}
\end{table}

All four points remain radiation dominated and reproduce the observed
BAU in the reference transport model. For $\Delta=(0.2,0.5,0.8,1.0)$,
the corresponding idealized three-year U-DECIGO SNRs are
\begin{align}
 M_{\rm PBH}=4.5\times10^5\,{\rm g}:&\quad (9.99,9.99,24.71,56.93),\label{eq:appF-snr-45}\\
 M_{\rm PBH}=6.0\times10^5\,{\rm g}:&\quad (14.98,14.96,36.96,85.02),\label{eq:appF-snr-60}\\
 M_{\rm PBH}=8.0\times10^5\,{\rm g}:&\quad (22.45,22.41,55.26,126.92),\label{eq:appF-snr-80}\\
 M_{\rm PBH}=1.0\times10^6\,{\rm g}:&\quad (30.72,30.64,75.48,173.14).\label{eq:appF-snr-100}
\end{align}
Increasing the PBH mass lowers the characteristic formation frequency
and correspondingly enhances the idealized SNR of the finite-width
infrared SIGW tail. The values are conditional forecasts using the
instrumental assumptions of Section~\ref{sec:GW}.

The four points in Table~\ref{tab:appF-original} vary the initial PBH mass while retuning $g_b$.
To test sensitivity beyond this one-dimensional family, we have repeated
the finite-temperature, continuous-evaporation transport calculation of
Section~\ref{sec:thermal}, varying the resonant splitting, the CP phases,
the initial PBH fraction, and selected masses and couplings. The return
damping matrix now includes the thermal $\psi$ Bose factor of Eq.~(\ref{eq:gammaR}),
and the central production coupling is $g_b=1.281\times10^{-5}$.
For each point below, $g_b\equiv g_{bY_1}=g_{bY_2}$ is the \emph{only} parameter
retuned after selecting the remaining inputs.  Parameters not specified
in the final column are fixed to Eq.~\eqref{eq:benchmark}.  All quoted BAU results
refer to the Bose-corrected reference transport system, not to a fully rederived
CP-odd inverse-reaction collision operator.

\begin{table}[t!]
\centering
\scriptsize
\resizebox{\textwidth}{!}{%
\begin{tabular}{lrrrrrrrl}
\hline\hline
Point & $M_{\rm PBH}/(10^5\,{\rm g})$ & $10^{12}\beta$
& $\Delta M$ [MeV] & $10^5g_b$ & $\Delta P_R(T=0)$
& $R_{\rm eq}^{\rm src}$ & $\max(\rho_{\rm PBH}/\rho_R)$
& Other changed inputs\\
\hline
B1  & 4.5  & 0.70 & 10 & 4.4714 & 0.0868 & 0.874 & 0.022 & ---\\
B2  & 8.0  & 1.20 & 27 & 0.5709 & 0.0916 & 0.419 & 0.067 & ---\\
B3  & 10.0 & 2.00 & 40 & 0.3198 & 0.0737 & 0.255 & 0.140 & ---\\
B5  & 6.0  & 1.00 & 12 & 1.1011 & 0.1695 & 0.660 & 0.042
    & $\phi_\kappa=0.75,\ \theta_y=2.35$\\
B7  & 6.0  & 2.00 & 15 & 2.0572 & 0.0994 & 0.670 & 0.084
    & $f=8.5\times10^{-5}$\\
B8  & 7.0  & 0.70 & 12 & 0.5951 & 0.0939 & 0.522 & 0.034
    & $f=6.0\times10^{-5}$\\
B9  & 6.0  & 0.50 & 18 & 1.8553 & 0.1005 & 0.664 & 0.021
    & $h=2.4\times10^{-10}$\\
B10 & 5.0  & 1.00 & 23 & 1.9336 & 0.0968 & 0.812 & 0.035
    & $h=5.2\times10^{-10}$\\
B11 & 6.5  & 1.00 & 18 & 1.1302 & 0.1005 & 0.602 & 0.046
    & $m_\phi=175\,{\rm GeV}$\\
B12 & 5.5  & 1.00 & 19 & 1.5903 & 0.0993 & 0.739 & 0.039
    & $m_\psi=9\,{\rm GeV}$\\
B13 & 6.0  & 1.20 & 12 & 1.1296 & 0.1220 & 0.666 & 0.051
    & $\kappa=0.030$\\
B14 & 6.0  & 0.90 & 28 & 1.4693 & 0.0714 & 0.667 & 0.038
    & $\kappa=0.055$\\
\hline\hline
\end{tabular}}
\caption{Additional benchmark points in the reference
continuous-evaporation transport model with a thermal $\psi$ population
in the return damping width. Each point has
$\eta_B^{\rm ref}=(8.720\pm0.001)\times10^{-11}$ after retuning $g_b$.
The quantity $R_{\rm eq}^{\rm src}$ is the abundance-only
source ratio defined in Appendix~\ref{app:G}; it is not
an independently corrected BAU prediction.}
\label{tab:appF-diversified}
\end{table}

In these points, the coherent return-probability difference is typically
$|\Delta P_R|=0.07$--$0.17$, even though the rare-production fraction
remains below $5.9\times10^{-3}$ in the implemented temperature range.
Thus, resonant enhancement of the conditional CP-odd branching probability
does not require the rare production reaction itself to equilibrate.
This differs from the loop-suppressed microscopic source studied in~\cite{Ambrosone:2021lsx}.

The examples span $\Delta M=10$--$40$ MeV, factors of four in $\beta$,
variations in both relevant CP phases, and changes in the loss,
return, and background-interaction parameters.  For all points,
$\max(\rho_{\rm PBH}/\rho_R)\lesssim0.141$, and the rare production rate
at $T=m_X$ remains below approximately $5\times10^{-5}H$.
The abundance-only ratios lie between $0.255$ and $0.874$,
illustrating order-unity, rather than exponential, depletion of the
integrated source in this fixed-kernel comparison.  The scan probes
robustness \emph{within} the resonant regime; it is not a symmetry-based
explanation of the small absolute splitting $\Delta M/M_Y$.

It is also useful to test nearby parameter choices \emph{without}
retuning $g_b$.  Fixing $g_b=1.281\times10^{-5}$ and all other inputs
to Eq.~\eqref{eq:benchmark}, we obtain the values illustrated in Table~\ref{tab:appF-untuned}. These untuned variations show that the Bose-corrected reference
transport result is not confined to an isolated numerical point, although the resonant
mass degeneracy itself remains a model-building assumption.  Changing
$M_{\rm PBH}$ or $\beta$ also changes the PBH-formation history;
therefore the idealized GW signal-to-noise ratios in Eqs.~\eqref{eq:appF-snr-45}--\eqref{eq:appF-snr-100}
are retained only for the corresponding original points.
\begin{table}[t!]
\centering
\small
\begin{tabular}{lrrrr}
\hline\hline
Varied input & Value 1 & $\eta_B^{\rm ref}/\eta_B^{\rm obs}$
& Value 2 & $\eta_B^{\rm ref}/\eta_B^{\rm obs}$\\
\hline
$\Delta M$ [MeV]   & $12$   & $0.838$ & $25$   & $1.109$\\
$\theta_y$          & $1.95$ & $0.561$ & $2.32$ & $1.288$\\
$\phi_\kappa$       & $0.90$ & $1.297$ & $1.16$ & $0.621$\\
$10^{12}\beta$      & $0.80$ & $0.806$ & $1.20$ & $1.192$\\
$g_b/g_b^{\rm new}$ & $0.90$ & $0.810$ & $1.10$ & $1.210$\\
\hline\hline
\end{tabular}
\caption{Untuned one-parameter variations about the Bose-corrected central
point. These are recalculations with the thermal $\psi$ return width
and $g_b=1.281\times10^{-5}$ unless $g_b$ itself is varied.}
\label{tab:appF-untuned}
\end{table}

\section{Appendix G: Impact of inverse reactions on the $X$ abundance}
\label{app:G}

In this appendix, we quantify the impact of the additional CP-even reactions
implied by the interaction Lagrangian in Eq.~\eqref{eq:int_lag} on the $X$ abundance during
the PBH-capture epoch.  Throughout this analysis we retain the PBH-capture
prescription, the temperature-dependent resonant CP-odd drag coefficient with the
thermal-$\psi$ return damping width, and the cosmological background
adopted in Section~\ref{sec:thermal}. 

In addition to the rare production reaction of Eq.~\eqref{eq:process1}, the interactions
$y_i\psi\overline{Y_i}X$ and $\kappa_{ir}S_r\overline{F_r}Y_i$
allow the reactions
\begin{equation}
 F_r+S_r\longleftrightarrow Y_i\longleftrightarrow X+\psi,
 \qquad X+\overline X\longleftrightarrow \psi+\psi^{(*)},
 \label{eq:appG-processes}
\end{equation}
where the final state of the last reaction is $\psi\psi$ for a real scalar
and $\psi\psi^\dagger$ for a complex scalar.  At the benchmark of
Eq.~\eqref{eq:benchmark}, the diagonal, zero-temperature partial widths are
\begin{equation}
 \Gamma_{Y_i\to F S}\simeq 1.623\times10^{-2}\,{\rm GeV},
 \qquad
 \Gamma_{Y_i\to X\psi}\simeq 6.939\times10^{-3}\,{\rm GeV}.
 \label{eq:appG-widths}
\end{equation}
A narrow-width estimate for the CP-even conversion reaction density is
\begin{equation}
 \gamma_{\rm conv}(T)\sim \sum_i n_{Y_i}^{\rm eq}(T)
 \frac{K_1(M_{Y_i}/T)}{K_2(M_{Y_i}/T)}
 \frac{\Gamma_{Y_i\to FS}\Gamma_{Y_i\to X\psi}}
      {\Gamma_{Y_i\to FS}+\Gamma_{Y_i\to X\psi}}.
 \label{eq:appG-reaction-density}
\end{equation}
This expression is used only to estimate the CP-even equilibration timescale;
in the nearly degenerate system the coherent amplitudes must be retained
when computing CP-odd reaction rates.

For the benchmark, the conversion and crossed-annihilation rates greatly
exceed the expansion rate during capture.  Representative results are
\begin{equation}
\begin{array}{c|cc}
 T\ [{\rm GeV}]& \Gamma_{\rm conv}/H & \Gamma_{X\overline X\to\psi\psi^{(*)}}/H\\ \hline
 130 & 1.46\times10^{11} & (1.25\text{--}1.41)\times10^7\\
 52.6 & 1.03\times10^{12} & (4.50\text{--}5.14)\times10^6
\end{array}
\label{eq:appG-rates}
\end{equation}
The ranges in the annihilation column correspond to the complex- and
real-$\psi$ conventions.  The rate estimates use on-shell CP-even widths, thermal $F_r,S_r$
daughters, and a Maxwell--Boltzmann treatment of the symmetric abundances.
They are intended to establish the rate hierarchy, rather than to replace
a full finite-density reaction-density calculation.
In particular, the small $\Gamma_{\rm prod}/H$ of the rare channel quoted
in Section~\ref{sec:thermal} does not imply that the entire $X,Y_i,\psi,F_r,S_r$ network
is out of equilibrium.

To assess the symmetric chemical relaxation, define
\begin{equation}
 n\equiv n_X+n_{\overline X},\qquad
 p\equiv n_\psi^{\rm tot},\qquad
 z_X\equiv\frac{n}{n^{\rm eq}},\qquad
 z_\psi\equiv\frac{p}{p^{\rm eq}}.
 \label{eq:appG-fugacities}
\end{equation}
In a CP-even quasistatic reduction of the short-lived $Y_i$ system,
with the thermal spectator daughters held in equilibrium, the collision
terms take the form
\begin{equation}
\begin{aligned}
 \dot n+3Hn & = \gamma_{\rm conv}(1-z_Xz_\psi)
   -2\gamma_{\rm ann}(z_X^2-z_\psi^2),\\
 \dot p+3Hp & = \gamma_{\rm conv}(1-z_Xz_\psi)
   +2\gamma_{\rm ann}(z_X^2-z_\psi^2),
\end{aligned}
\label{eq:appG-symmetric-evolution}
\end{equation}
where $\gamma_{\rm ann}\equiv(n_X^{\rm eq})^2
\langle\sigma_{X\overline X\to\psi\psi^{(*)}}v\rangle$.
Both reactions together imply $z_Xz_\psi\simeq1$ and
$z_X^2\simeq z_\psi^2$; hence, for positive abundances,
\begin{equation}
 z_X\simeq z_\psi\simeq1.
 \label{eq:appG-fast-equilibrium}
\end{equation}
We have integrated Eq.~\eqref{eq:appG-symmetric-evolution} from
$x=m_X/T=0.1$ through the PBH-capture interval, on the same continuous
PBH-evaporation background as the reference calculation.  In this CP-even
Maxwell--Boltzmann reduction, the $X$ abundance follows its equilibrium
value to better than $10^{-6}$ during the capture epoch.  This statement
concerns symmetric chemical equilibration; it does not fix the CP-odd
charge-transport source.

For consistency with the revised thermal treatment, the reference CP-odd
source in the following diagnostic is evaluated using
$\Gamma_R(T)=[1-f_F(E_X,T)+f_B(E_\psi,T)]\Gamma_R(0)$, with
$E_\psi\simeq35.37\,{\rm GeV}$ and the retuned central coupling
$g_b=1.281\times10^{-5}$. These choices modify the effective
source used for the abundance comparison; they do not constitute a full
inverse-reaction derivation of the CP-odd momentum-transfer kernel. The equilibrium yield for one of $X$ and $\overline X$, with $g_X=2$ and
$g_{*s}=100$, is
\begin{equation}
 Y_X^{\rm eq}(x)\equiv\frac{n_X^{\rm eq}}s
 =\frac{45g_X}{4\pi^4g_{*s}}x^2K_2(x).
 \label{eq:appG-equilibrium-yield}
\end{equation}
Compared with the $X$ abundance of the reference transport solution,
the equilibrium abundance during capture is
\begin{equation}
\begin{array}{c|cccc}
 T\ [{\rm GeV}]&130&100&70&52.6\\ \hline
 Y_X^{\rm eq}/Y_X^{\rm ref}&1.000&0.887&0.681&0.495
\end{array}.
\label{eq:appG-abundance-ratio}
\end{equation}
The PBHs evaporate at $T\simeq52.6\,{\rm GeV}$, corresponding to
$x\simeq2.47$; consequently, they stop capturing before the $X$
population becomes exponentially small.

There is a useful integral check independent of the separate spectator
and $X$-decay transfer terms.  Adding the evolution equations for $n_-$, $n_B$, and $q_B$ in Section~\ref{sec:thermal} gives
\begin{equation}
 \dot Q_B+3HQ_B
 =\Gamma_{\rm abs}\bigl(\delta_{\rm drag}n_+-n_-\bigr),
 \qquad Q_B\equiv n_-+n_B+q_B.
 \label{eq:appG-charge-identity}
\end{equation}
For vanishing initial $Q_B$ and final comoving entropy $S_f=s_fa_f^3$,
this implies
\begin{equation}
 \eta_B^{\rm final}=
 \frac1{S_f}\int du\,\frac{a^3\Gamma_{\rm abs}}{H}
 \bigl(\delta_{\rm drag}n_+-n_-\bigr),
 \qquad u\equiv\ln a.
 \label{eq:appG-source-integral}
\end{equation}
To isolate the abundance effect, we replace $n_+^{\rm ref}$ by
$n_+^{\rm eq}$ only in the positive source term, leaving
$\Gamma_{\rm abs}$, $\delta_{\rm drag}$, $H$, $S_f$, and the reference
$n_-$ term unchanged.  We find
\begin{equation}
 R_{\rm eq}^{\rm src}\equiv
 \frac{\displaystyle\int du\,\frac{a^3\Gamma_{\rm abs}}H
                \delta_{\rm drag}n_+^{\rm eq}}
      {\displaystyle\int du\,\frac{a^3\Gamma_{\rm abs}}H
                \delta_{\rm drag}n_+^{\rm ref}}
 =0.665.
 \label{eq:appG-source-ratio}
\end{equation}
The source-only correction is therefore a reduction of $33.5\%$, rather
than a many-orders-of-magnitude suppression.  With the reference
backreaction term held fixed, the corresponding \emph{conditional}
late-time estimate is $\eta_B^{\rm sens}\simeq5.79\times10^{-11}$.
Within this fixed-kernel test, retuning the production coupling from
$g_b=1.281\times10^{-5}$ to approximately $1.573\times10^{-5}$ restores
$\eta_B^{\rm sens}\simeq8.72\times10^{-11}$.




\end{document}